\documentclass[twocolumn,superscriptaddress,
 amsmath,amssymb,
 aps,
 prl,
]{revtex4-1}
\usepackage{dcolumn}
\usepackage{bm}
\usepackage{graphicx}
\usepackage{physics}
\usepackage{comment}
\usepackage[caption=false]{subfig}
\usepackage{color}

\newcommand{\bea}{\begin{eqnarray}}
\newcommand{\eea}{\end{eqnarray}}

\newcommand{\blue}{\textcolor{black}}

\begin{document}

\title{Dynamically emergent correlations between particles in a switching harmonic trap}
\author{Marco Biroli}
\affiliation{LPTMS, CNRS, Univ.  Paris-Sud,  Universit\'e Paris-Saclay,  91405 Orsay,  France}
\author{Manas Kulkarni}
\affiliation{ICTS, Tata Institute of Fundamental Research, 560089 Bengaluru, India}
\author{Satya N. Majumdar}
\affiliation{LPTMS, CNRS, Univ.  Paris-Sud,  Universit\'e Paris-Saclay,  91405 Orsay,  France}
\author{Gr\'egory Schehr}
\affiliation{Sorbonne Universit\'e, Laboratoire de Physique Th\'eorique et Hautes Energies, CNRS UMR 7589, 4 Place Jussieu, 75252 Paris Cedex 05, France}


\begin{abstract}
We study a one dimensional gas of $N$ noninteracting diffusing particles in a harmonic trap, whose stiffness switches between two values $\mu_1$ and $\mu_2$ with constant rates $r_1$ and $r_2$ respectively. Despite the absence of direct interaction between the particles,  we show that strong correlations between them emerge in the stationary state at long times, induced purely by the dynamics itself. We compute exactly the joint distribution of the positions of the particles in the stationary state, which allows us to compute several physical observables analytically. In particular, we show that the extreme value statistics (EVS), i.e., the distribution of the position of the rightmost particle has a nontrivial shape in the large $N$ limit. The scaling function characterizing this EVS has a finite support with a tunable shape (by varying the parameters).
Remarkably, this scaling function turns out to be universal. First, it also describes the distribution of the position of the $k$-th rightmost particle in a $1d$ trap. Moreover, the distribution of the position of the particle farthest from the center of the harmonic trap in $d$ dimensions is also described by the same scaling function for all $d \geq 1$. Numerical simulations are in excellent agreement with our analytical predictions. 
\end{abstract}

\maketitle

Stochastic resetting (SR) has emerged as a major area of research in statistical physics with multidisciplinary applications across
diverse fields, such as search algorithms in computer science, foraging processes in ecology, reaction-diffusion processes 
in chemistry, and transcription processes in biology~\cite{EMS20,PKR22,GJ22}. SR simply means interrupting the natural dynamics of a system at random times
and instantaneously restarting the process either from its initial configuration or more generally from any pre-decided state. The interval between
two successive resettings is typically Poissonian, though other protocols such as periodic resetting have also been studied. 
One of the main effects of SR is that the resetting moves violate detailed balance and drive the system to a non-equilibrium stationary state (NESS)~\cite{EM11a,EM11b}. Characterising such a NESS and its possible spatial structure has generated a lot of interest, both theoretically (for reviews see~\cite{EMS20,PKR22,GJ22}) and experimentally~\cite{Roichman20,BBPMC_20,FBPCM_21}.
One of the simplest theoretical models corresponds to a single diffusing particle in $d$ dimensions and subjected to SR with a constant rate $r$ (i.e., Poissonian resetting)~\cite{EM11a,EM11b}. In this case, the position distribution becomes time independent at long times and has a nontrivial non-Gaussian shape. This result has been verified experimentally in optical traps setups~\cite{Roichman20}. Subsequently, several other models of single particle noisy dynamics subject to stochastic resetting have been studied theoretically~\cite{Reuveni14,Boyer14,Reuveni15,Majumdar15b,PKE_16,Reu_16,MV_16,PR_17,BEM_17,EM18,CS_18,Boyer18,PKR_19,Maso19c,Pal20,BS_20,BS_2_20,Bres_20,Pinsky_20,BMS_22,Mori22,SBS22,NG16,BMS_23,BRR20,MOK23,BM23}. 

\begin{figure}
\centering
\includegraphics[width = \linewidth]{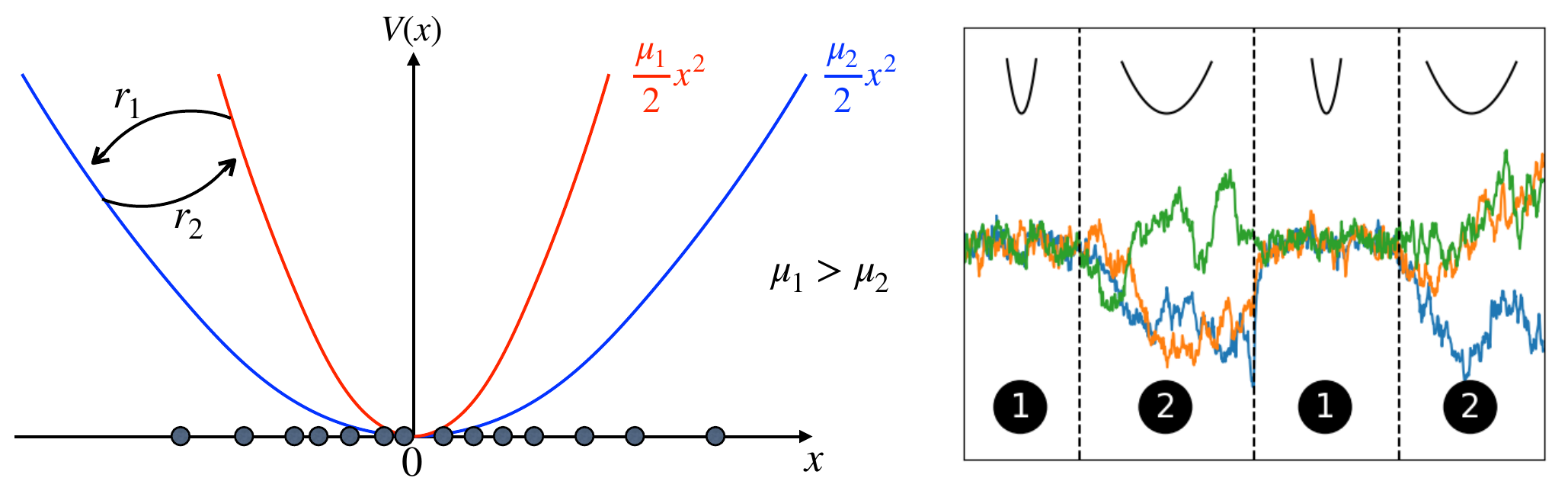}
\caption{The left panel illustrates the setup where $N$ particles are confined in a harmonic potential $V(x)$, whose stiffness alternates between $\mu_1$ and $\mu_2$ with rates $r_1$ (from $\mu_1$ to $\mu_2$) and $r_2$ (from $\mu_2$ to $\mu_1$) respectively. On the right, we show 
the schematic trajectories for $N = 3$ diffusing particles. The switching times are shown by dashed vertical black lines. } \label{fig:ex-rho}
\end{figure}

The stochastic resetting for single particle systems discussed above can be easily generalized to many-body systems. 
In this case, the whole configuration of the system (i.e., all the degrees of freedom) is reset instantaneously to a pre-decided
configuration at random times with rate $r$. This leads to a many-body NESS with interesting spatial structures that have been observed
in a number of systems, such as fluctuating interfaces~\cite{GMS14},  symmetric exclusion process~\cite{BKP19}, the Ising model~\cite{MMS20}, etc. 
Recently, a very simple
model of $N$ noninteracting Brownian motions in one dimension, subjected to simultaneous resetting to their initial positions with rate 
$r$, was introduced~\cite{BLMS_23}. Remarkably, even though the particles are noninteracting in this model, the simultaneous resetting generates an effective all-to-all 
attractive interaction between these particles that persists even at long times in the NESS. This model demonstrated an important phenomenon,
namely the {\it emergence of strong collective correlations} in the steady state of a many-body system, where the interactions between constituents are not built-in but instead emerge from the dynamics itself. 

One of the shortcomings of these theoretical models, either for single or multi-particle system, is the assumption of {\it instantaneous} resetting~\cite{Reu_16,EM18_ref,BS_20,BS_2_20,Sunil23}.
While this assumption makes the problem simpler and
easy to implement in both numerical simulations and theoretical analysis, it is not very realistic experimentally. 
For example, in the optical trap experiments of a single diffusing particle with SR, the typical protocol consists of alternative intervals of free diffusion and confined motions~\cite{BBPMC_20,FBPCM_21}. During the free period, the particle is allowed to diffuse freely in the absence of an optical trap. At the end of this period a harmonic trap is switched on and the particle is thermally equilibrated in the trap. Once the particle has equilibrated, the trap is switched off and a new period of free motion starts. In the confined phase, no measurement is performed, since this protocol was designed to mimic the instantaneous resetting move~\cite{BBPMC_20,FBPCM_21}. One may naturally wonder what happens if one does not wait till the full equilibration in the confined phase but instead switches off the trap at
a random time, e.g., distributed exponentially.

This leads to a more realistic and general protocol where the particle moves in a harmonic trap whose 
stiffness switches intermittently between $\mu_1$ and $\mu_2$ (with $\mu_1 > \mu_2$ without any loss of generality). The stiffness changes from $\mu_1$ to $\mu_2$ with rate $r_1$ and reciprocally with rate $r_2$ from $\mu_2$ to $\mu_1$ (see Fig. \ref{fig:ex-rho} for an illustration). In the limit $\mu_1 \to \infty$, $\mu_2 \to 0$ and $r_1 \to \infty$, this general protocol reduces to the standard model of diffusion under SR to the origin. The limit $\mu_1 \to \infty$ and $\mu_2 \to 0$ ensures resetting of a diffusing particle to the origin, while the limit $r_1 \to \infty$ guarantees that once it is reset to the origin, it immediately restarts, thus realising the instantaneous resetting. For a {\it single} particle undergoing this switching intermittent potential, the 
resulting position distribution in the NESS has been studied only recently~\cite{MBMS_20,GPKP_20,SDN_21,XZMD_22,GP_22,ACB_22,MBM_22}. In this paper, our goal is to study $N$ independent particles undergoing this switching intermittent protocol. One of our main results is to show that, indeed, the switching dynamics between two stiffnesses of the trap drives the system into a NESS with strong collective correlations that emerge purely out of the dynamics. Thus the emergence of strong correlations without direct interaction is a robust phenomenon and is not just an artefact of instantaneous resetting.

\begin{figure*}
\centering
\includegraphics[width=0.32\textwidth]{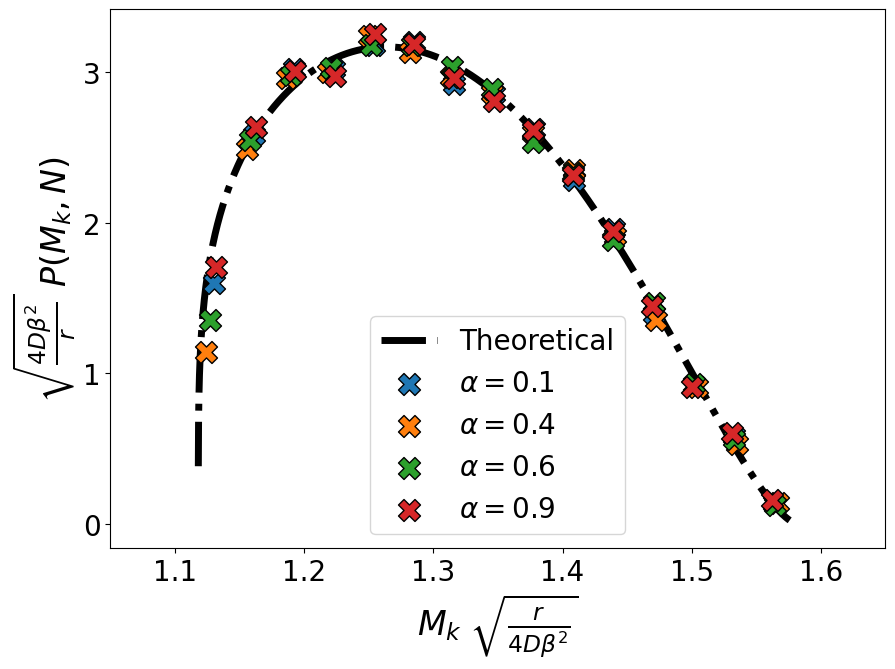} 
\hfill
\includegraphics[width=0.32\textwidth]{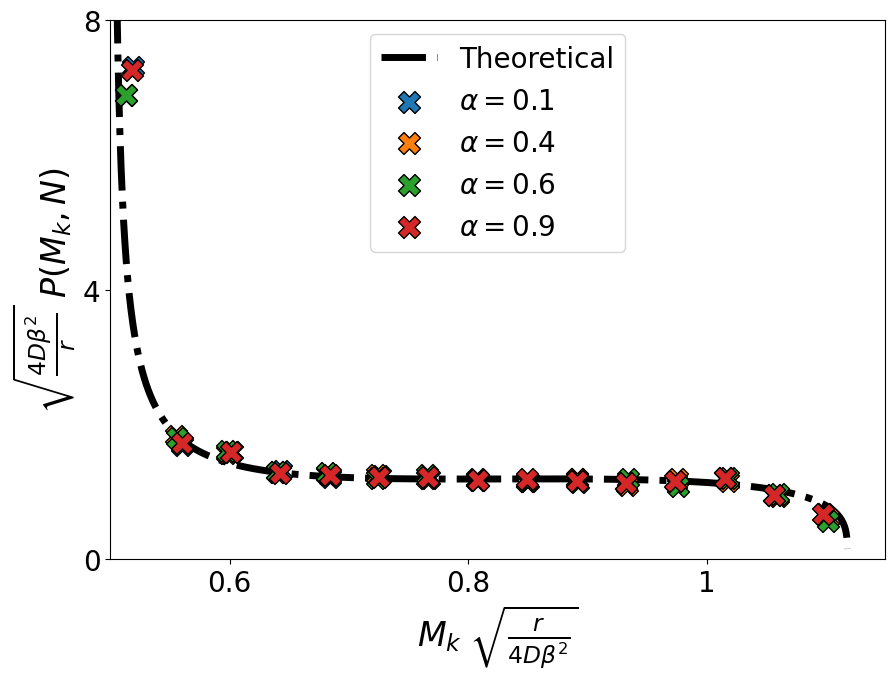} 
\hfill
\includegraphics[width=0.32\textwidth]{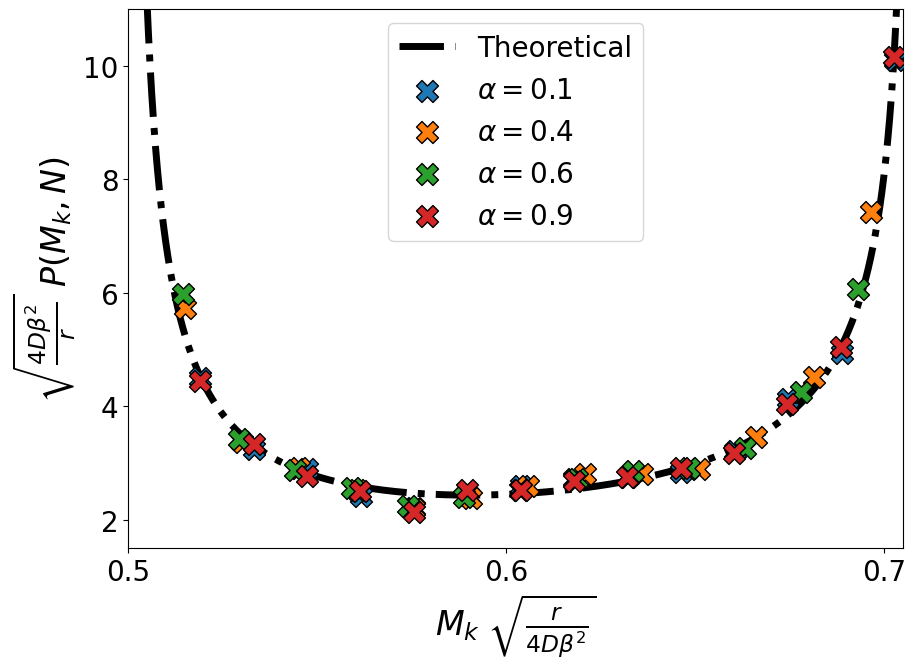} 
\caption{Scaling collapse of the distribution of the $k$-th maximum as in Eq.~(\ref{Mk}) for different values of $\alpha = k/N$ and different values of the parameters. We set $r_1 = r_2 = 1$, $D=1$, $N = 10^{6}$ and vary $\mu_1$ and $\mu_2$. From left to right we used respectively $\mu_1 = 0.4, \, \mu_2 = 0.2$ then $\mu_1 = 2, \, \mu_2 = 0.4$ and finally $\mu_1 = 2,\, \mu_2 = 1$. The dashed black line corresponds to the theoretical prediction and the symbols are the numerical results. Different colors correspond to different values of $\alpha$. The numerical results were obtained by sampling $10^{5}$ examples directly from the NESS distribution given in Eq. (\ref{ness}).} \label{fig:Mk}
\end{figure*}

Let us first summarize our main results. For $N$ independent particles on the line driven by this switching intermittent protocol, we first provide a  
complete characterisation of the NESS, i.e., the exact computation of the joint distribution of the positions of the particles. This allows us to compute the spatial correlations in the NESS, as well as several other physical observables, such as the average density, the distribution of the position of the rightmost particle in the gas (extreme value statistics), the spacing distribution between particles and the full counting statistics (FCS), i.e., the statistics of the number of particles in a given interval. These observables have been calculated recently for large $N$ in the limit of instantaneous resetting \cite{BLMS_23} but, here, we show that these asymptotic results get drastically modified under this intermittent switching protocol. In particular, we find a surprising result for the extreme value statistics (EVS), i.e., the distribution of the position $M_1$ of the rightmost particle. We show that in the large $N$ limit, $M_1$ typically scales as $\sqrt{\ln N}$ and its probability distribution function (PDF) takes the scaling form
\begin{equation} \label{scaling_M1}
{\rm Prob.}\left( M_1 = w, N\right) \approx \sqrt{\frac{r_H}{4D\, \ln N}} f\left( w \sqrt{\frac{r_H}{4D\, \ln N}}\right) 
\end{equation}  
where $r_H = 2/(1/r_1+1/r_2)$ is the harmonic mean of the switching rates and the scaling function $f(z)$ has a nontrivial shape 
supported over a {\it finite interval} $\sqrt{r_H/(2 \mu_1)} \leq z \leq \sqrt{r_H/(2 \mu_2)}$ [see Eqs. \eqref{f} and \eqref{Mk} and Fig. \ref{fig:Mk}], even though the average density is supported over the full line (see Fig. \ref{fig:ex-rho}). By tuning the parameters $r_1, r_2, \mu_1, \mu_2$, the shape of this PDF changes drastically as seen in Fig. \ref{fig:Mk}. This is remarkable since in all the known examples of EVS in 
uncorrelated~\cite{G_58,A_78,LLR_82,D_85,W_88,ABN_92,ND_03,FC_15} or correlated~\cite{TW_94,DM_01,CD_01,MK_03,MC_04,MC_05,SM_06,BC_06}
 systems (for a recent review see~\cite{MPS_20}), including the instantaneous resetting case discussed above, the limiting distribution of the maximum is always supported over an unbounded interval (infinite or semi-infinite). The emergence of a finite support with a tunable shape for the EVS is thus a strong signature of the non-instantaneous nature of this switching protocol. In addition to having a finite support, we find that the scaling function $f(z)$ in Eq. (\ref{scaling_M1}) is surprisingly robust and universal: it also describes the scaling of the $k$-th maximum in $d=1$ as well as the distribution of the distance of the farthest particle from the center of a $d$-dimensional harmonic trap. In the rest of the paper, we present only the computation of the joint distribution and the EVS in Eq. (\ref{scaling_M1}). The computations of the other observables mentioned above are provided in the Supp. Mat. \cite{SM}.

\vspace{0.2cm}

{\emph{The Model.}} We consider $N$ independent Brownian particles on a line, all starting at the origin which feel a potential that switches between 
$V_1(x) = \mu_1 x^2/2$ and $V_2(x) = \mu_2 x^2/2$,  
with Poissonian rate $r_1$ (from $\mu_1$ to $\mu_2$) and rate $r_2$ (from $\mu_2$ to $\mu_1$). Hence, the duration $\tau$ of the time intervals  between successive switches is distributed via ${\rm Prob.}[\tau] = r_i e^{-r_i\tau}$, where $r_i$ is $r_1$ or $r_2$. Moreover, the intervals are statistically independent. In each phase the positions $\{ x_i\}$ evolve as independent Ornstein-Uhlenbeck processes~\cite{UO_30}
\bea \label{dynamic}
\dv{x_i}{t} = - \mu_k x_i + \sqrt{2 D} \eta_i(t) \;,
\eea
where $\mu_k = \mu_1$ or $\mu_2$ depending on the phase, $D$ is the diffusion constant and $\eta_i(t)$ is a zero-mean Gaussian white noise with a correlator $\langle \eta_i(t) \eta_j(t') \rangle = \delta_{ij} \delta(t - t')$. Let $P_1(\vec{x}, t)$ (resp. $P_2$) denote the joint PDF of the particles being at $\vec{x}=(x_1, \cdots, x_N)$ at time $t$ and that the system is in phase 1 (resp. phase 2). From Eq. (\ref{dynamic}), they evolve by the coupled Fokker-Planck equations
\bea
\pdv{P_1}{t} &= \sum\limits_{i = 1}^N \left[D \pdv[2]{P_1}{x_i} + \mu_1 \pdv{}{x_i} \left(x_i P_1 \right) \right] - r_1 P_1 + r_2 P_2 \label{fp-p1}\\
\pdv{P_2}{t} &= \sum\limits_{i = 1}^N \left[ D \pdv[2]{P_2}{x_i} + \mu_2 \pdv{}{x_i} \left( x_i P_2 \right) \right] - r_2 P_2 + r_1 P_1 \label{fp-p2}
\eea
with the initial conditions
\bea \label{init-fp}
P_1(\vec{x}, t = 0) = \frac{1}{2}\delta(\vec{x}) \mbox{~~and~~} P_2(\vec{x}, t = 0) = \frac{1}{2}\delta(\vec{x}) \;,
\eea
where we assumed that, initially, both phases occur equally likely. Hence the joint PDF of the positions only is given by $P(\vec{x}, t) = P_1(\vec{x}, t) + P_2(\vec{x}, t)$. The first terms on the right hand side of Eqs. (\ref{fp-p1}) and (\ref{fp-p2}) represent diffusion and advection in a harmonic potential, while the last two terms represent the loss and gain due to the switching between potentials, with rates $r_1$ and $r_2$ respectively.  

To solve this pair of Fokker-Planck equations, it is convenient to work in the Fourier space where we define $\tilde P_n(\vec{k},t) = \int_{-\infty}^{+\infty} \dd \vec{x} \; e^{i \vec{k} \dot  \vec{x}} P_n(\vec{x},t)$, with $n=1,2$. In the steady state, setting $\partial_t \tilde P_n = 0$, Eqs. (\ref{fp-p1}) and (\ref{fp-p2}) 
in the Fourier space reduce to
\bea
\left(D \sum\limits_{i = 1}^N k_i^2 + r_1\right) \tilde{P_1} + \mu_1 \sum\limits_{i = 1}^N k_i \pdv{\tilde{P_1}}{k_i} &= r_2 \tilde{P_2}\label{fourier-fp-p1}\\
\left(D \sum\limits_{i = 1}^N k_i^2 + r_2\right) \tilde{P_2} + \mu_2 \sum_{i = 1}^N k_i \pdv{\tilde{P_2}}{k_i} &= r_1 \tilde{P_1} \label{fourier-fp-p2}\;,
\eea
with initial conditions $\tilde{P_n}(\vec{k} = 0) = 1/2$. Notice that Eqs (\ref{fourier-fp-p1})-(\ref{fourier-fp-p2}) are spherically symmetric. It is therefore much easier to move to hyper-spherical coordinates where $k = \sqrt{\sum_{i = 1}^N k_i^2}$ is the distance to the origin and $\theta_i$, for $i = 1, \cdots, N-1$ are the different angular coordinates. Then Eqs. (\ref{fourier-fp-p1})-(\ref{fourier-fp-p2}) simplify to
\bea
\left[(D k^2 + r_1) + \mu_1 k \partial_k \right]\tilde{P_1} &= r_2 \tilde{P_2} \label{spherical-fourier-fp-p1}\\
\left[(D k^2 + r_2) + \mu_2 k \partial_k\right] \tilde{P_2} &= r_1 \tilde{P_1} \label{spherical-fourier-fp-p2} \;.
\eea 
Notice that by permuting the indices $1 \leftrightarrow 2$ in Eq. (\ref{spherical-fourier-fp-p1}) leads to Eq. (\ref{spherical-fourier-fp-p2}). Hence, we can solve only for $\tilde{P_1}$ and the solution for $\tilde{P_2}$ will follow by permuting the indices. By eliminating $\tilde P_2$ between Eqs. (\ref{spherical-fourier-fp-p1}) and (\ref{spherical-fourier-fp-p2})  we get an ordinary second order differential equation for $\tilde P_1$ (respectively $\tilde P_2$). Solving this ordinary differential equations with appropriate boundary conditions (see Supp. Mat. for details) we obtain 
\begin{equation}
\tilde{P_1}(k) = \frac{r_2\,e^{- \frac{D k^2}{2 \mu_1}}}{r_1+r_2} M\left( R_1; 1 + R_1 + R_2; \frac{D k^2(\mu_2 - \mu_1)}{2 \mu_1 \mu_2} \right), \label{P1-k}
\end{equation}
where $R_1 = r_1/(2\mu_1)$, $R_2 = r_2/(2 \mu_2)$ and $M(a;b;z)$ is the Kummer's function \cite{NIST}. Similarly, one can obtain $\tilde P_2(\vec{k})$ just by exchanging $\mu_1 \leftrightarrow \mu_2$ and $r_1 \leftrightarrow r_2$. To reveal the spatial correlations in the NESS, it is useful to invert this Fourier transform, which is not easy. However, fortunately, one can make use of a convenient integral representation~\cite{NIST} 
\begin{equation} 
M(a ; b ; z) = \frac{\Gamma(b)}{\Gamma(a) \Gamma(b - a)} \int_0^1 \dd u\; e^{z u} u^{a-1} (1 - u)^{b - a - 1} \;, \label{M-integral}
\end{equation}
where $\Gamma(x)$ is the Gamma function. Using Eq. (\ref{M-integral}) in Eq. (\ref{P1-k}) and inverting the Fourier transform we obtain an expression for $P_1(\vec{x})$ and \blue{similarly for} $P_2(\vec{x})$. Adding them gives the joint PDF in the NESS~\cite{SM}
\begin{eqnarray} \label{ness}
P^{\rm st} (\vec{x})= \int_0^{1} \dd u \; h(u) \prod_{i = 1}^N p(x_i | u) \;,
\end{eqnarray}
where 
\bea \label{h}
h(u) = \frac{c\,r_H}{4} u^{R_1 - 1} (1 - u)^{R_2 - 1} \left[ \frac{1 - u}{\mu_1} + \frac{u}{\mu_2} \right] 
\eea
with $c = \Gamma(R_1 + R_2 + 1) / ( \Gamma(R_1 + 1) \Gamma(R_2 + 1) )$ and $r_H = 2 \, r_1 r_2/(r_1+r_2)$. The function $p(x | u) = e^{- \frac{x^2}{2 V(u)}}/\sqrt{2 \pi V(u)}$ is a pure Gaussian with zero mean and 
variance $V(u) = D \left( \frac{u}{\mu_2} + \frac{1 - u}{\mu_1} \right)$.
This fully characterizes the joint PDF of the positions in the NESS. Note that $h(u)$ is normalised to unity, i.e., $\int_0^1 h(u) \dd u = 1$. Thus one can interpret Eq. (\ref{ness}) as the joint distribution of $N$ i.i.d. Gaussian variables with zero mean and a common variance $V(u)$ parametrised by $u$, which itself is a random variable distributed via the PDF $h(u)$. There is indeed a nice physical meaning of this random variable $u$. If the particle was entirely in phase $2$, its stationary distribution would be a Gaussian (the Gibbs state) with a variance $D/\mu_2$. In contrast, if it was in phase $1$, it will again be a Gaussian with a variance $D/\mu_1$. Hence from the formula $V(u) = D \left( \frac{u}{\mu_2} + \frac{1 - u}{\mu_1} \right)$, one sees that $0 \leq u \leq 1$ can be interpreted as the effective fraction of time that each particle spends in phase $2$. This can be put on a more rigorous footing by using the so-called Kesten variables as shown in the Supp. Mat. \cite{SM}. For simplicity, we will henceforth 
set $r_1=r_2=r$ and the results for general $r_1 \neq r_2$ are given in the Supp. Mat.~\cite{SM}.

We note that the joint PDF in Eq.~(\ref{ness}) does not factorise, indicating the presence of correlations in the NESS. One can easily calculate the two-point correlation function from Eq. (\ref{ness}) using the fact that, for a fixed $u$, they are i.i.d. variables. 
The natural correlator $\langle x_i x_j\rangle - \langle x_i\rangle \langle x_j\rangle $ for $i \neq j$ vanishes identically since $p(x_i \vert u)$ is Gaussian and hence symmetric in $x_i$. The first nonzero correlator for $i \neq j$ is given by
\begin{align} \label{correlator}
\langle x_i^2 x_j^2 \rangle &- \langle x_i^2 \rangle \langle x_j^2 \rangle = \nonumber\\ 
&\frac{D^2}{r^2} \frac{(R_1 - R_2)^2 ( 2 + 3 R_1 + 3 R_2 + 4 R_1 R_2)}{(1 + R_1 + R_2)^2 (2 + R_1 + R_2)} \;,
\end{align}
where we recall that $R_1 = r/(2 \mu_1)$ and $R_2= r/(2 \mu_2)$. The positive value of this correlator indicates that there are effective all-to-all attractive correlations between the particles in the NESS. These correlations are not built-in but get generated by the switching dynamics of the potential, which all the particles share together. This makes the particles strongly correlated in the NESS. Despite such strong correlations, the structure of the joint PDF in Eq. (\ref{ness}) allows us to compute several physical observables exactly, such as the average density, the EVS, the distribution of the spacings between particles and also the FCS. The reason for the solvability can be traced back to Eq. (\ref{ness}) where one can first fix $u$ and compute the observables for $N$ independent variables, each distributed via $p(x\vert u)$ where $u$ is just a fixed parameter and then average over $u$ drawn from the PDF $h(u)$ in Eq. (\ref{h}). For i.i.d. variables, this computation is rather standard. This solvable structure holds more generally for any conditionally independent and identically distributed  (c.i.i.d.) variables, as studied recently in Ref.~\cite{BLMS_23-2}. Here, the c.i.i.d. structure emerges from the basic dynamics of the system and thus provides a natural physical example of such systems. The computation of these physical observables are provided in details in the Supp. Mat. \cite{SM} and here we focus only on the EVS. This is because the EVS of strongly correlated variables is known to be a very hard problem and there are only few cases where it can be derived analytically. Our model provides not only a solvable example of EVS in a strongly correlated system, but also the distribution of the EVS turns out to be rather surprising as discussed below.

To compute the EVS, we start from the joint PDF in Eq. (\ref{ness}). We first fix $u$ and compute the EVS of $N$ i.i.d. Gaussian random variables of zero mean and variance $V(u)$. It is well known~\blue{\cite{MPS_20}} that, for large $N$, the maximum $M_1$ of such i.i.d. Gaussian variables behaves almost deterministically as $M_1 \approx \sqrt{2 V(u)\, \ln N}$, with fluctuations around it of order $1/\sqrt{\ln N}$. It turns out that, to leading order for large $N$, one can approximate this distribution by a delta function, namely $P(M_1 = w\vert u) \approx \delta(w-\sqrt{ 2V(u)\, \ln N})$. Finally, averaging over $u$ we get
\begin{equation}\label{PDF_EVS}
\hspace*{-0.25cm}P(M_1 = w,N) \approx \int_0^1 \dd u \, h(u) \, \delta\left(w-\sqrt{ 2V(u)\, \ln N}\right) 
\end{equation}
where $V(u) = D(u/\mu_2 + (1-u)/\mu_1)$ and $h(u)$ is given in Eq. (\ref{h}). Performing this integral explicitly~\cite{SM}, we get the scaling form in Eq. (\ref{scaling_M1}) where the scaling function $f(z)$ has a nontrivial shape given by 
\begin{align} \label{f}
f(z) = \frac{c\, R_1^{R_1 - 1} R_2^{R_2 - 1}}{(R_2 - R_1)^{R_1 + R_2 - 1}} &|z|^3 \nonumber\\
\left( 1 - \frac{z^2}{R_2} \right)^{R_2 - 1}& \left( \frac{z^2}{R_1} - 1 \right)^{R_1 - 1} \;,
\end{align}
with $\sqrt{R_1} \leq z \leq \sqrt{R_2}$. As mentioned earlier, an EVS scaling function with a finite support is rather surprising because the average density is spread over the full real line~\cite{SM}. Moreover the shape of the scaling function $f(z)$ can be tuned by varying the parameters $R_1$ and $R_2$. At both edges of the support $f(z)$ can either diverge, go to a nonzero constant or vanish, depending on $R_1, R_2$. The scaling function $f(z)$ also turns out to be universal in the following sense. If one calculates the distribution of the $k$-th maximum (order statistics), one finds a scaling form 
\bea \label{Mk}
{\rm Prob.}[M_k = w,N] \approx \sqrt{\frac{r_H}{4 D \beta^2}} \; f\left( w\sqrt{\frac{r_H}{4 D \beta^2}} \right) \;,
\eea
where $\beta = {\rm erfc}^{-1}(2 k/N)$, but the scaling function $f(z)$ is independent of $k$ and has the same expression as in Eq.~(\ref{f}). Here ${\rm erfc}(z) = 2/\sqrt{\pi} \int_{z}^\infty e^{-y^2}\,\dd y$. In Fig. \ref{fig:Mk}, we verify this scaling form by collapsing data for different $\alpha = k/N$ and for different values of $R_1$ and $R_2$. The numerical results are in excellent agreement with our theoretical predictions. Furthermore, one can easily generalise our results to a harmonic trap in $d$ dimensions~\blue{\cite{SM}}. Following exactly the same analysis as in the $d=1$ case above, one can also compute the
distribution of the distance of the farthest particle from the center of the trap and we find the remarkable result that it is again described by Eq. (\ref{scaling_M1}) with the same scaling function $f(z)$ given in Eq. (\ref{f}). Thus the scaling function $f(z)$ is extremely robust and ``super-universal'', in the sense that it neither depends on $k$ in $d=1$ and nor on the dimension $d$ itself.

To summarize, we have completely characterised the nonequilibrium stationary state of $N$ Brownian particles in a harmonic trap in an experimentally realistic protocol where the stiffness of the trap switches between two values at constant rate. The strong correlations between the positions of the particles in the stationary state emerge from the dynamics itself and are not built-in. The exact joint distribution of the particle positions allows us to compute several physical observables analytically. In particular, we have shown that the EVS is characterized by a nontrivial scaling function which has a finite support and a tunable shape. Moreover, the scaling function of the EVS is universal in the sense that it also describes the limiting distribution of the $k$-th maximum in $d=1$ as well as the distribution of the distance of the particle farthest from the center of the harmonic trap in $d$-dimensions \cite{SM}. It would be interesting if our predictions can be verified experimentally and also to investigate \blue{the NESS in} non-harmonic traps. 

\acknowledgments {\it Acknowledgments}. MK and SNM would like to thank the Isaac
Newton Institute for Mathematical Sciences, Cambridge, for support and
hospitality during the programme {\it New statistical physics in living
matter: non equilibrium states under adaptive control} where work on
this paper was undertaken. This work was supported by EPSRC Grant
Number EP/R014604/1. MK would like to acknowledge support from the CEFIPRA Project 6004-1, SERB Matrics Grant (MTR/2019/001101) and SERB VAJRA faculty scheme (VJR/2019/000079). MK acknowledges support from the Department of Atomic Energy, Government of India, under Project No. RTI4001.

\end{document}


\title{Dynamically emergent correlations between particles in a switching harmonic trap: Supplementary material}
\author{Marco Biroli}
\affiliation{LPTMS, CNRS, Univ.  Paris-Sud,  Universit\'e Paris-Saclay,  91405 Orsay,  France}
\author{Manas Kulkarni}
\affiliation{ICTS, Tata Institute of Fundamental Research, 560089 Bengaluru, India}
\author{Satya N. Majumdar}
\affiliation{LPTMS, CNRS, Univ.  Paris-Sud,  Universit\'e Paris-Saclay,  91405 Orsay,  France}
\author{Gr\'egory Schehr}
\affiliation{Sorbonne Universit\'e, Laboratoire de Physique Th\'eorique et Hautes Energies, CNRS UMR 7589, 4 Place Jussieu, 75252 Paris Cedex 05, France}


    
{
\let\clearpage\relax  
\let\hfill\relax           
\maketitle
}

\tableofcontents


\section{Derivation of the joint distribution of the positions in the stationary state for general $r_1 \neq r_2$}\label{sec_deriv}

In this Section we provide the details of the computation of the joint probability distribution function (JPDF) of the positions of $N$ particles in the NESS
for \blue{arbitrary} switching rates $r_1$ and  $r_2$. Here, $r_1$ represents the rate at which the potential switches from $\mu_1 x^2/2$ to $\mu_2 x^2/2$ and $r_2$ represents the reverse rate (see Fig. 1 of the main text). Then, the Fokker-Planck equations as presented in the Letter read
\bea
\pdv{P_1}{t} &= \sum\limits_{i = 1}^N D \pdv[2]{P_1}{x_i} + \mu_1 \pdv{}{x_i} \left(x_i P_1 \right) - r_1 P_1 + r_2 P_2 \label{fp-p1}\\
\pdv{P_2}{t} &= \sum\limits_{i = 1}^N D \pdv[2]{P_2}{x_i} + \mu_2 \pdv{}{x_i} \left( x_i P_2 \right) - r_2 P_2 + r_1 P_1 \label{fp-p2}\;,
\eea
where $P_1(\vec{x}, t)$ (resp. $P_2(\vec{x}, t)$) is the joint probability of the particles being at positions $\vec{x}$ at time $t$ \blue{and in phase 1} (resp. in phase 2). We \blue{consider symmetric} initial conditions
\bea \label{init-fp}
P_1(\vec{x}, t = 0) = \frac{1}{2} \delta(\vec{x}) \mbox{~~and~~} P_2(\vec{x}, t = 0) = \frac{1}{2} \delta(\vec{x}) \;.
\eea
The JPDF $P(\vec{x}, t)$ \blue{of the positions $\vec x$ at time $t$, regardless of the phase of the
potential, is then given by}
\bea \label{P-P1-P2}
P(\vec{x}, t) = P_1(\vec{x}, t) + P_2(\vec{x}, t) \;.
\eea
We now proceed with the computation of the JPDF in the NESS. As discussed in the Letter, it is convenient to work in the Fourier space. Hence we define 
\bea
\tilde{P}_n(\vec{k}, t) = \int_{-\infty}^{+\infty} \dd x_1 \cdots \int_{-\infty}^{+\infty} \dd x_N \; e^{i \sum_{j = 1}^N k_j x_j} P_n(\vec{x},t) \;,
\eea
with $n=1,2$. In the long time limit, the system reaches a stationary state which is obtained by setting the left hand side of Eqs. (\ref{fp-p1}) and \blue{(\ref{fp-p2})} to zero. This gives, in the Fourier space
\begin{align}
&\left( D \sum_{i = 1}^N k_i^2 + r_1 \right) \tilde{P}_1 + \mu_1 \sum_{i = 1}^N k_i \pdv{\tilde{P}_1}{k_i} = r_2 \tilde{P}_2 \label{pde-ns-p1} \\
&\left( D \sum_{i = 1}^N k_i^2 + r_2 \right) \tilde{P}_2 + \mu_2 \sum_{i = 1}^N k_i \pdv{\tilde{P}_2}{k_i} = r_1 \tilde{P}_1 \;, \label{pde-ns-p2}
\end{align}
where $\tilde P_n(\vec{k})$ denote the stationary JPDF in the Fourier space. Given that \blue{both Eqs. (\ref{pde-ns-p1}) and (\ref{pde-ns-p2})} are spherically symmetric we can considerably simplify them by changing variables to $k = \sqrt{\sum_{i = 1}^N k_i^2}$. Performing this substitution in Eq. (\ref{pde-ns-p1}) and Eq. (\ref{pde-ns-p2}) and making use of the spherical symmetry yields
\begin{align}
&\left[ \left(D k^2 + r_1 \right) + \mu_1 k  \frac{d}{dk} \right] \tilde{P}_1 = {\cal L}_1 \tilde{P}_1 = r_2 \tilde{P}_2 \label{pde-p1}\\
&\left[ \left(D k^2 + r_2 \right) + \mu_2 k \frac{d}{dk} \right] \tilde{P}_2 = {\cal L}_2 \tilde{P}_2 = r_1 \tilde{P}_1 \label{pde-p2}\;,
\end{align}
where ${\cal L}_n = \left[ \left(D k^2 + r_n \right) + \mu_n k \frac{d}{dk}\right]$, with $n=1,2$. Note that, by setting $k=0$ we get the relation
$r_1 \tilde P_1(k=0) = r_2 \tilde P_2(k=0)$. In addition, we have $\tilde P_1(k=0)+\tilde P_2(k=0) = 1$, due to normalization. Solving this sets the boundary condition at $k=0$, namely
\begin{eqnarray} \label{bc_Fourier}
\tilde P_1(k=0) = \frac{r_2}{r_1+r_2} \quad {\rm and} \quad \tilde P_2(k=0) = \frac{r_1}{r_1+r_2} \;.
\end{eqnarray}
To solve Eqs. (\ref{pde-p1}) and (\ref{pde-p2}) we proceed as follows. 
We first act on Eq. (\ref{pde-p1}) with ${\cal L}_2$ and on Eq. (\ref{pde-p2}) with ${\cal L}_1$. These allow us to decouple these coupled ordinary differential equations (ODEs) 
\begin{align}
&\left[ \left(D k^2 + r_2 \right) + \mu_2 k \frac{d}{dk} \right]\left[ \left(D k^2 + r_1 \right) + \mu_1 k \frac{d}{dk} \right] \tilde{P}_1 = r_1 r_2 \tilde{P}_1 \label{ode-p1}\\
&\left[ \left(D k^2 + r_1 \right) + \mu_1 k \frac{d}{dk} \right]\left[ \left(D k^2 + r_2 \right) + \mu_2 k \frac{d}{dk} \right] \tilde{P}_2 = r_1 r_2 \tilde{P}_2 \label{ode-p2}\;.
\end{align}
Notice that switching the indices $1 \leftrightarrow 2$ transforms Eq. (\ref{ode-p1}) into Eq. (\ref{ode-p2}). Hence, we can restrict ourselves to solving the ODE in Eq. (\ref{ode-p1}), the solution of Eq. (\ref{ode-p2}) will be obtained by permuting the indices $1 \leftrightarrow 2$ in the solution of Eq. (\ref{ode-p1}). Solving (which can be done with Mathematica) the ODE in Eq. (\ref{ode-p1}) yields the most general solution
\bea \label{M-U-res}
\tilde{P}_1(k) =  e^{-\frac{D k^2}{2 \mu_1}} \;\left[ A_1 M\left( R_1; 1 + R_1 + R_2; - \frac{D k^2 (\mu_1 - \mu_2)}{2 \mu_1 \mu_2} \right) + B_1 U\left( R_1; 1 + R_1 + R_2; - \frac{D k^2 (\mu_1 - \mu_2)}{2 \mu_1 \mu_2} \right) \right]\;,
\eea
where $A_1, B_1$ are arbitrary constants and we denote
\bea \label{R1-R2}
R_1 = \frac{r_1}{2 \mu_1} \mbox{~~and~~} R_2 = \frac{r_2}{2 \mu_2} \;.
\eea
Here $M(a;b;z)$ is the Kummer's function defined by the power series \cite{NIST}
\bea \label{Kummer}
M(a;b;z) = 1 + \frac{a}{b}z + \frac{a(a+1)}{b(b+1)} \frac{z^2}{2!} + \cdots \;. 
\eea
and $U(a;b;z)$ is the confluent hypergeometric $U$ function. 
To fix the constants $A_1$ and $B_1$ we will use the boundary conditions in Eq. (\ref{bc_Fourier}). We use the small argument asymptotics of the hypergeometric functions, namely  $M(a;b;z) \to 1$ and $U(a;b;z) \sim z^{1-b}$ as $z \to 0$. Taking the $k\to 0$ limit in Eq. (\ref{M-U-res}), one sees that the second term diverges as $k^{-2(R_1+R_2)}$ as $k \to 0$. However, from Eq. (\ref{bc_Fourier}), we see that $\tilde P_1(k=0) = {r_2}/{(r_1+r_2)}$. Hence we must have $B_1 = 0$. Taking the limit $k \to 0$ in Eq. (\ref{M-U-res}) then fixes $A_1 = r_2/(r_1+r_2)$. Similarly, one can write down the solution for $\tilde P_2(k)$ by exchanging the indices $1$ and $2$. This gives
\bea \label{tilde-p1}
\tilde{P}_1(k) = \frac{r_2}{r_1 + r_2} e^{-\frac{D k^2}{2 \mu_1}} M \left( R_1; 1 + R_1 + R_2; - \frac{D k^2 (\mu_1 - \mu_2)}{2 \mu_1 \mu_2} \right)
\eea
and
\bea \label{tilde-p2}
\tilde{P}_2(k) = \frac{r_1}{r_1 + r_2} e^{-\frac{D k^2}{2 \mu_2}} M \left( R_2; 1 + R_1 + R_2; - \frac{D k^2 (\mu_2 - \mu_1)}{2 \mu_1 \mu_2} \right) \;.
\eea

While these results are exact in the Fourier space, it is not easy to extract the spatial correlations between the particles from these Fourier representations. For this purpose, it would be useful to invert this Fourier transform if possible. Fortunately, it \blue{turns out} that there is a very nice integral representation of the Kummer's function which reads~\cite{NIST}
\bea \label{M-integral}
M(a; b; z) = \frac{\Gamma(b)}{\Gamma(a)\Gamma(b - a)} \int_0^1 \dd u \; e^{z u} u^{a - 1} (1 - u)^{b - a - 1} \;. 
\eea
Using Eq. (\ref{M-integral}) we can re-express Eq. (\ref{tilde-p1}) as
\bea \label{tilde-p1-int}
\tilde{P}_1(k) = \frac{r_2}{r_1 + r_2} \frac{\Gamma(1 + R_1 + R_2)}{\Gamma(R_1) \Gamma(1 + R_2)} \int_0^1 \dd u \; u^{R_1 - 1} (1 - u)^{R_2} e^{-k^2\left(\frac{D(u \mu_1 + (1 - u)\mu_2)}{2 \mu_1 \mu_2}\right)} \;.
\eea
Under this form, one can now easily invert the Fourier transform by using the identity 
\bea \label{gaussian-invert}
 \int_{-\infty}^{+\infty} \frac{\dd k_1}{2 \pi} e^{-i k_1 x_1} \cdots \int_{-\infty}^{+\infty} \frac{\dd k_N}{2 \pi} e^{-i k_N x_N} \;  e^{-k^2 a} = \prod_{i = 1}^N \frac{1}{\sqrt{4 \pi a}} e^{-\frac{x_i^2}{4 a}} \;,
\eea
provided $a > 0$. Using the \blue{result (\ref{gaussian-invert})} in Eq. (\ref{tilde-p1-int}) we obtain the real space representation of the JPDF
\bea \label{p1-x}
P_1(x_1, \ldots, x_N) = \frac{r_2}{r_1 + r_2} \frac{R_1\,\Gamma(1 + R_1 + R_2)}{\Gamma(R_1 + 1) \Gamma(1 + R_2)} \int_0^1 \dd u \;  u^{R_1 - 1} (1 - u)^{R_2} \prod_{i = 1}^N \frac{1}{\sqrt{2 \pi V(u)}} e^{-\frac{x_i^2}{2 V(u)}} \;, \
\eea
where
\bea \label{def_V_of_u}
V(u) = D\,\left( \frac{u}{\mu_2} + \frac{1 - u}{\mu_1} \right) \;.
\eea
An identical computation for $\tilde{P_2}(k)$ yields
\bea \label{p2-x-inverted}
P_2(x_1, \ldots, x_N) = \frac{r_1}{r_1 + r_2} \frac{ R_2\,\Gamma(1 + R_1 + R_2) }{\Gamma(R_1 + 1) \Gamma(R_2 + 1)} \int_0^1 \dd u \;  u^{R_2 - 1} (1 - u)^{R_1} \prod_{i = 1}^N \frac{1}{\sqrt{2 \pi V(u)}} e^{-\frac{x_i^2}{2 V(u)}} \;.
\eea
In order to make Eq. (\ref{p2-x-inverted}) as similar to Eq. (\ref{p1-x}) as possible and therefore allowing for a more compact result it is convenient to make the change of variable $u \to 1 - u$ in Eq. (\ref{p2-x-inverted}) which yields
\bea \label{p2-x}
P_2(x_1, \ldots, x_N) = \frac{r_1}{r_1 + r_2} \frac{ R_2\,\Gamma(1 + R_1 + R_2) }{\Gamma(R_1 + 1) \Gamma(R_2 + 1)} \int_0^1 \dd u \; R_2 u^{R_1} (1 - u)^{R_2 - 1} \prod_{i = 1}^N \frac{1}{\sqrt{2 \pi V(u)}} e^{-\frac{x_i^2}{2 V(u)}} \;.
\eea
\blue{Adding (\ref{p1-x}) and (\ref{p2-x}), we get}
%
\begin{equation} \label{P-x}
P(\vec{x}) = \frac{r_1 r_2}{2(r_1 + r_2)}\,\frac{\Gamma(1 + R_1 + R_2)}{\Gamma(R_1 + 1)\Gamma(R_2 + 1)}  \int_0^1 \dd u \; u^{R_1 - 1} (1 - u)^{R_2 - 1} \left[ \frac{u}{\mu_2} + \frac{1 - u}{\mu_1}\right] \prod_{i = 1}^N \frac{1}{\sqrt{2 \pi V(u)}} e^{-\frac{x_i^2}{2 V(u)}} \;.
\end{equation}
This result can be written in a compact form 
\bea \label{Pst_sm}
P^{\rm st}(\vec x) = P(\vec{x}) = \int_0^1 {\rm d}u\, h(u) \prod_{i=1}^N p(x_i \vert u) \;,
\eea
where 
\bea \label{h_u_sm}
h(u) = \frac{c\,r_H}{4} u^{R_1 - 1} (1 - u)^{R_2 - 1} \left[ \frac{1 - u}{\mu_1} + \frac{u}{\mu_2} \right] 
\eea
with $c = \Gamma(R_1 + R_2 + 1) / ( \Gamma(R_1 + 1) \Gamma(R_2 + 1) )$ and $r_H = 2 \, r_1 r_2/(r_1+r_2)$. Here the function $p(x | u)$ is a pure Gaussian with zero mean and 
variance $V(u)$ given in Eq. (\ref{def_V_of_u}), i.e.,  
\bea \label{p}
p(x | u) = \frac{1}{\sqrt{2 \pi V(u)}} \; e^{- \frac{x^2}{2 V(u)}} \;.
\eea
One can check that $h(u)$ is normalized to unity, i.e.,
\bea \label{norm_h}
\int_0^1 {\rm d}u\, h(u) = 1 \;.
\eea
Since $h(u) \geq 0$ and \blue{is} normalised to unity, it can be interpreted as a PDF of a random variable $u$. Thus one can interpret Eq. (\ref{Pst_sm}) as the joint distribution of $N$ i.i.d. Gaussian variables with zero mean and a common variance $V(u)$ parametrised by $u$, which itself is a random variable distributed via the PDF $h(u)$. There is indeed a nice physical meaning of this random variable $u$. If the particle was entirely in phase $2$, its stationary distribution would be a Gaussian (the Gibbs state) with a variance $D/\mu_2$. In contrast, if it was in phase $1$, it will again be a Gaussian with a variance $D/\mu_1$. Hence from the formula for \blue{$V(u)$ in Eq. (\ref{def_V_of_u})}, one sees that $0 \leq u \leq 1$ can be interpreted as the effective fraction of time the particle spends in phase $2$. This can be put on a more rigorous footing by using the so-called Kesten \blue{recursion relations} as shown in details in the next section.

\section{Kesten Approach}

As mentioned in the Letter, we are considering $N$ particles at positions $\vec{x} = x_1, \cdots, x_N$ which diffuse independently within a potential \blue{that} switches from $V_1(x) = \frac{1}{2} \mu_1 x^2$ and $V_2(x) = \frac{1}{2} \mu_2 x^2$ with Poissonian rates $r_1$ and $r_2$, i.e. with rate $r_1$ the system will switch from $V_1(x)$ to $V_2(x)$ and respectively with rate $r_2$ it will switch back from $V_2(x)$ to $V_1(x)$. Hence, the duration $\tau$ of the time intervals \blue{between} successive switches is distributed as
\bea \label{p-tau}
{\rm Prob.}[\tau] = r_i e^{- r_i \tau} \;,
\eea
where $r_i$ is $r_1$ or $r_2$ according to which potential is on during the interval. \blue{Furthermore, all intervals are distributed independently from each other. We denote by $\{\tau_i\}$ all the successive intervals}. Without loss of generality assume that $V_1(x)$ \blue{is} on during the odd intervals $\{\tau_1, \tau_3, \cdots\}$ and respectively $V_2(x)$ \blue{is} on during the even intervals $\{\tau_2, \tau_4, \cdots\}$. During the odd intervals the equation of motion is
\bea \label{dynamic1}
\dv{x_i}{\tau} = -\mu_1 x_i + \sqrt{2 D} \eta_i(\tau) \;,
\eea
and respectively during the even intervals the equation of motion is
\bea \label{dynamic2}
\dv{x_i}{\tau} = -\mu_2 x_i + \sqrt{2 D} \eta_i(\tau) \;.
\eea
As stated in the Letter, $\eta_i(\tau)$ is a Gaussian white noise such that
\bea \label{eta}
\langle \eta_i(\tau) \rangle = 0 \mbox{~~and~~} \langle \eta_i(\tau) \eta_j(\tau') \rangle = \delta(\tau - \tau')\delta_{ij} \;.
\eea
We saw in the Letter that making use of the spherical symmetry the problem reduces to a one-dimensional problem on $x = \left(\sum_{i = 1}^N x_i^2\right)^{1/2}$. For simplicity, we will restrict ourselves below to the one-particle case, but the multi-particle case follows the same derivation.


For a fixed choice of random intervals $\{\tau_i\}$ the process $x(\tau)$ is a Gaussian process since \blue{Eqs. (\ref{dynamic1}) and (\ref{dynamic2})} are linear evolution equations. Hence the probability distribution $P(x, \tau | \{\tau_i\})$ of the system being at $x$ at time $\tau$ knowing the random intervals $\{\tau_i\}$ is given by
\bea \label{p-x-tau}
P(x, \tau | \{\tau_i\} ) = \frac{1}{\sqrt{2 \pi \; V(\tau, \{\tau_i\})}} \exp( - \frac{x^2}{2 \; V(\tau, \{\tau_i\})}) \;,
\eea
where $V(\tau, \{\tau_i\})$ is the variance at time $\tau$ given the $\{\tau_i\}$'s. In the long time limit \blue{the system reaches a steady state}. Hence when $\tau \to +\infty$ then $V(\tau, \{\tau_i\}) \to V(\{\tau_i\})$. Therefore the distribution $P(x | \{\tau_i\})$ in the steady state is given by
\bea \label{p-x}
P(x | \{\tau_i\} ) = \frac{1}{\sqrt{2 \pi\; V(\{\tau_i\})}} \exp( - \frac{x^2}{2 \; V(\{\tau_i\})}) \;.
\eea
If we now average over all possible realizations of the $\{\tau_i\}$'s it will give us the stationary distribution $P^{\rm st}(x)$,
\bea \label{p-st}
P^{\rm st}(x) = \int_0^{+\infty} \dd V \; {\rm Prob.}[V] \frac{1}{\sqrt{2 \pi  V}} \exp( - \frac{x^2}{2  V})\;,
\eea
\blue{where $V$ stands for $V(\{\tau_i\})$ averaged over the interval lengths $\tau_i$'s.} We recognize here a form similar to Eq. (\ref{P-x}). The goal now is to find the distribution ${\rm Prob.}[V]$ which is induced by the random variables $\{\tau_1, \tau_2, \cdots\}$, this should allow us to recover Eq. (\ref{P-x}). To do so we will proceed recursively. \blue{Let $x_n$ denote the position of the particle $x(\tau)$ 
at the end of the $n$-th interval, i.e., $x_n= x\left(\tau= \sum_{i=1}^n \tau_i\right)$.}
Furthermore, we denote by $V_n = \langle x_n^2 \rangle$ the variance at the end of the $n$-th interval. If $n$ is odd, then the potential $V_1(x)$ was on during that interval. Hence 
\bea \label{xodd-xeven}
x_{2n + 1} = x_{2n} e^{-\mu_1 \tau_{2n + 1}} + \sqrt{2 D} \; e^{-\mu_1 \tau_{2n + 1}} \int_0^{\tau_{2n + 1}} \eta(\tau) e^{\mu_1 \tau} \dd \tau \;,
\eea 
and consequently
\bea \label{Vodd-Veven}
V_{2n + 1} = V_{2n} e^{-2 \mu_1 \tau_{2n + 1}} + \frac{D}{\mu_1}\left(1 - e^{-2 \mu_1 \tau_{2n + 1}}\right)
\eea
Inversely, if $n$ is even then the potential $V_2(x)$ was on during that interval. Hence
\bea \label{xeven-xodd}
x_{2n} = x_{2n-1} e^{-\mu_2 \tau_{2n}} + \sqrt{2 D} \; e^{-\mu_2 \tau_{2n}} \int_0^{\tau_{2n}} \eta(\tau) e^{\mu_2 \tau} \dd \tau \;,
\eea
and consequently
\bea \label{Veven-Vodd}
V_{2n} = V_{2n-1} e^{-2 \mu_2 \tau_{2n}} + \frac{D}{\mu_2}\left(1 - e^{-2 \mu_2 \tau_{2n}}\right) \;.
\eea
Here $\tau_n$'s are random variables drawn from two exponential distributions $r_1 e^{-r_1 \tau}$ and  $r_2 e^{-r_2 \tau}$ alternatively. Thus the recursion relations satisfied by the $V_n$'s involve random variables. Such linear recursion relations with random coefficients are known as ``Kesten recursion relations'' and they have been studied in different contexts such as in probability theory and disordered systems~\cite{Kesten73,KKS75,DH83,KS84,CLNP85,G91,BDMZ13,GBL21,mathis}.   

From (\ref{Vodd-Veven}) and (\ref{Veven-Vodd}), we see that $V_n$ fluctuates between $D/\mu_1$ and $D/\mu_2$ in the large $n$ limit. It is then convenient to define a new variable 
\bea \label{def_un}
u_n = \frac{V_n - \frac{D}{\mu_1}}{\frac{D}{\mu_2} - \frac{D}{\mu_1}} \;.
\eea
Therefore $u_n$ lies in the interval $[0,1]$ in the $n \to \infty$ \blue{limit} (we recall that $\mu_1>\mu_2$).  Applying the reparametrization in Eq. (\ref{def_un}) to Eqs. (\ref{Vodd-Veven}) and (\ref{Veven-Vodd}) we get the recursion relations
\bea \label{t-rec}
u_{2n + 1} = u_{2n} e^{-2 \mu_1 \tau_{2n + 1}} \mbox{~~and~~} 1 - u_{2n} = (1 - u_{2n - 1}) e^{-2 \mu_2 \tau_n} \;.
\eea 
It is convenient to define $z_n = e^{-2 \mu_i \tau_n}$ where $\mu_i = \mu_1$ if $n$ is odd and $\mu_i = \mu_2$ otherwise. From Eq. (\ref{p-tau}) we get
\bea \label{z}
{\rm Prob.}[z_{2n + 1} = z] = R_1 z^{R_1 - 1} \mbox{~~and~~} {\rm Prob.}[z_n = z] = R_2 z^{R_2 - 1}\;, \qquad {\rm with} \; 0 \leq z \leq 1 \;,
\eea
where $R_i = r_i/(2 \mu_i)$. In the $n \to +\infty$ limit we expect to reach a steady state. Hence we expect that $u_{2n + 1} \to u_{\rm odd}$ and $u_{2n} \to u_{\rm even}$ as $n \to +\infty$, and from Eqs. (\ref{t-rec}) and (\ref{z}) we know that
\bea \label{t-st-coupled}
u_{\rm odd} = u_{\rm even} z_1 \mbox{~~and~~} 1 - u_{\rm even} = (1 - u_{\rm odd}) z_2 \;.
\eea
Let $P_{\rm even}(u)$ and $P_{\rm odd}(u)$ denote respectively the stationary distribution of $u_{\rm even}$ and $u_{\rm odd}$ in the limit $n \to \infty$. Then, from Eq. (\ref{t-st-coupled}) we have 
\bea
P_{\rm odd}(u) = \int_0^1 \dd u' \int_0^1 \dd z \; P_{\rm even}(u') R_1 z^{R_1 - 1} \delta(u - u' z) = \int_u^1 \dd z\; P_{\rm even}\left(\frac{u}{z}\right) R_1 z^{R_1 - 1} \;.
\eea
Making a change of variable to $y = u/z$ we get
\bea
P_{\rm odd}(u) = R_1 u^{R_1 - 1} \int_u^{1} \frac{P_{\rm even}(y)}{y^{R_1}} \dd y\;.
\eea
Taking a derivative with respect to $u$, one gets
\bea \label{ode-odd}
\dv{}{u} \left[ \frac{1}{u^{R_1 - 1}} P_{\rm odd}(u) \right] = - \frac{R_1}{u^{R_1}} P_{\rm even}(u) \;.
\eea
A similar derivation for $P_{\rm even}(u)$ leads to
\bea \label{ode-even}
\dv{}{(1 - u)} \left[ \frac{1}{(1 - u)^{R_2 -1}} P_{\rm even}(1 - u) \right] = \frac{-R_2}{(1 - u)^{R_2}} P_{\rm odd}(1 - u) \;.
\eea
Given the ODEs in Eqs. (\ref{ode-odd}) and (\ref{ode-even}) one can easily check that the solutions are given by 
\bea
P_{\rm odd}(u) = \blue{c\, R_1 \, u^{R_1 - 1} (1 - u)^{R_2} }\mbox{~~and~~} P_{\rm even}(u) = \blue{c \, R_2 \, u^{R_1} (1 - u)^{R_2 - 1}} \;,
\eea
where $c$ is an arbitrary constant, yet to be fixed. To fix this constant, we proceed as follows. From Eq. (\ref{bc_Fourier}), we know that, in the stationary state, the potential itself is in phase $1$ (with stiffness $\mu_1$) with probability $r_2/(r_1+r_2)$ and is in phase $2$ (with stiffness $\mu_2$) with the complementary probability $r_1/(r_1+r_2)$. Therefore, the PDF $P_{\rm odd}(u)$ will occur with probability $r_2/(r_1+r_2)$ and $P_{\rm even}(u)$ will occur with probability $r_1/(r_1+r_2)$. Hence the full PDF of the random variable $u$ in the stationary state is given by
\bea \label{ht-prelim}
h(u) = \frac{r_2 P_{\rm odd}(u) + r_1 P_{\rm even}(u)}{r_1 + r_2} = \frac{\blue{c\,R_1\, R_2}}{r_1 + r_2} u^{R_1-1} (1 - u)^{R_2-1} \left[ \frac{1 - u}{R_2} r_1 + \frac{ u }{R_1} r_2 \right] \;.
\eea
The normalization condition $\int_0^{1} \dd u \; h(u) = 1$ then fixes the constant $c$ to be
\bea \label{c}
\blue{c = \frac{\Gamma(R_1 + R_2 + 1)}{\Gamma(R_1+1) \Gamma(R_2+1)}} \;.
\eea
Placing Eq. (\ref{c}) back in Eq. (\ref{ht-prelim}) we obtain
\bea \label{h}
h(u) = \frac{r_1 r_2}{2(r_1 + r_2)} \frac{\Gamma(R_1 + R_2 + 1)}{\Gamma(R_1 + 1) \Gamma(R_2 + 1)} \; u^{R_1 - 1} (1 - u)^{R_2 - 1} \left[\frac{u}{\mu_2} +\frac{1 - u}{\mu_1}\right] \;,
\eea
recovering exactly Eq. (\ref{h_u_sm}). This Kesten approach thus shows clearly that the random variable $u$ has the physical interpretation of the fraction of time the particle spends in phase 2.

\vspace{0.5cm}

\section{Observables in a one-dimensional switching harmonic trap}

In this Section, we will derive in detail the statistics of all the observables in the NESS, as discussed in the main text. This includes the average density, the first non-trivial correlator, the EVS and the order statistics, the gap statistics and also the full counting statistics, i.e. the number of particles in an interval $[-L, L]$ around the origin, which we did not discuss in the main text. All the derivations follow from the JPDF in the NESS given in Eq. (\ref{Pst_sm}), which we recall for convenience   
\bea \label{jpdf}
P^{\rm st}(\vec{x}) = \int_0^1 \dd u\; h(u) \prod_{k = 1}^N p(x_k | u) \;,
\eea
where 
\bea \label{h-p}
h(u) = \frac{c \, r_H}{4} u^{R_1 - 1} (1 - u)^{R_2 - 1} \left[\frac{u}{\mu_2} + \frac{1 - u}{\mu_1} \right] \mbox{~~and~~} p(x | u) = \sqrt{\frac{1}{2 \pi V(u)}} \exp( - \frac{x^2}{2 V(u)} )  \;,
\eea
with $V(u)$ given in Eq. (\ref{def_V_of_u}) and the constants are 
\bea \label{const_sm}
c = \frac{\Gamma(R_1 + R_2 + 1)}{\Gamma(R_1 + 1)\Gamma(R_2 + 1)}\;, \quad R_1 = \frac{r_1}{2 \mu_1} \;, \quad R_2 = \frac{r_2}{2 \mu_2} \quad {\rm and} \quad r_H = 2 \frac{r_1 r_2}{r_1+r_2} \;.
\eea

\subsection{Average density}

\vspace{0.5cm}

\begin{figure}
\centering
\includegraphics[width=0.7\textwidth]{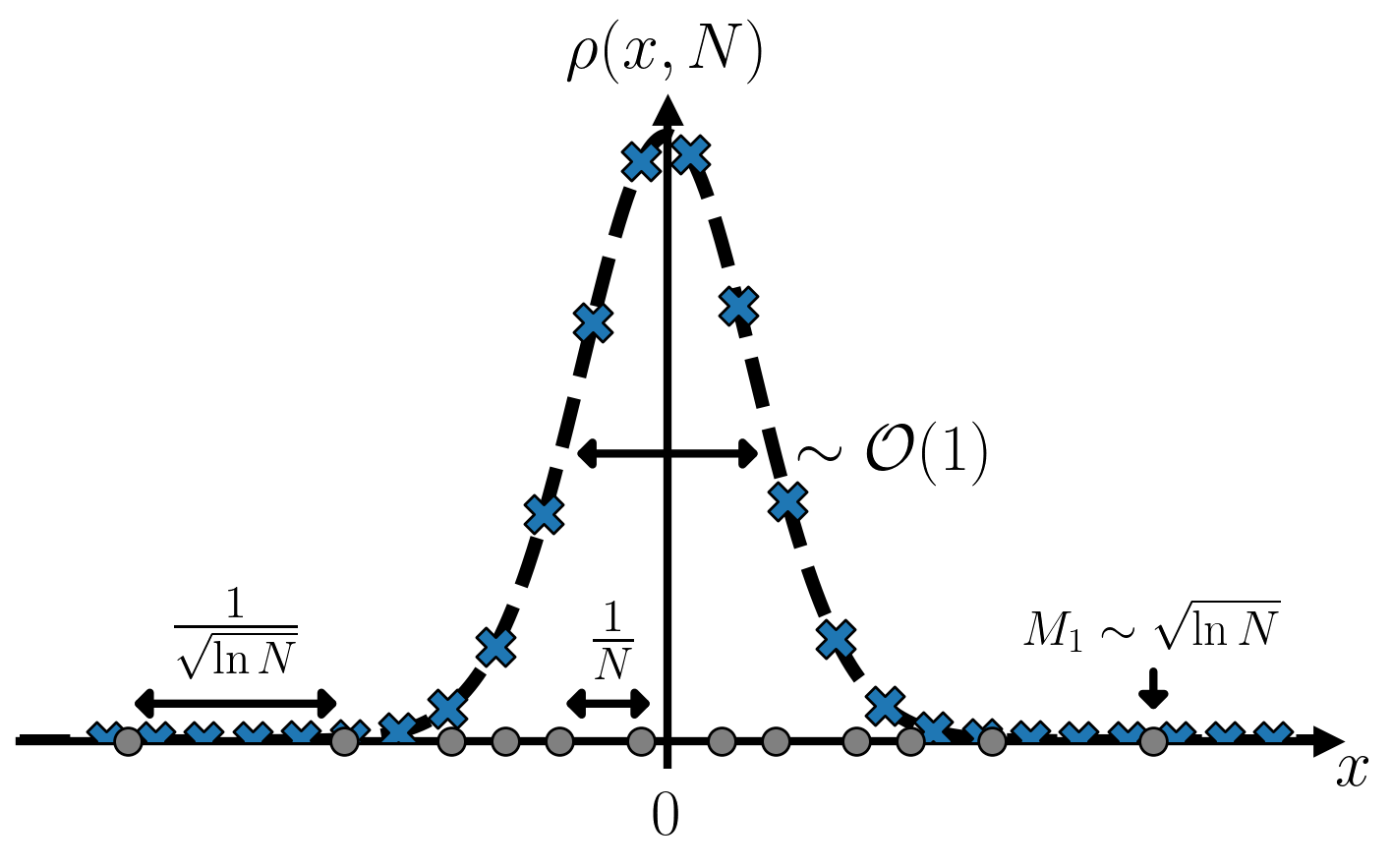}
\caption{A sketch of a typical configuration of the system. The solid blue line shows the average density $\rho(x, N)$. The positions of the particles in a typical sample is shown schematically on the line with most particles living over a distance $\sqrt{2 D /r_H}$ around the origin, where $r_H$ denotes the harmonic mean of $r_1$ and $r_2$. The typical spacing in the bulk scales like $1/N$ while it is of order $1/\sqrt{\ln N}$ near the extreme edges of the sample. The typical position of the rightmost particle $M_1$ scales like $\sqrt{\ln N}$.} \label{fig-supp-mat}
\end{figure}  

We start with the most basic observable, namely the average density of the gas defined by
\bea \label{def_rho}
\rho(x, N) = \frac{1}{N}\left\langle \sum_{i = 1}^N \delta(x - x_i) \right\rangle \;,
\eea
which is normalized to unity $\int_{-\infty}^{\infty} \dd x\; \rho(x,N) = 1$. Here $\langle \cdots \rangle$ means an average over the JPDF in Eq. (\ref{jpdf}). 
From the expression of the JPDF, it is clear that $\rho(x,N)$ is the one-point marginal distribution, i.e.,  
\begin{equation} \label{rho-x}
\rho(x, N) = \int_{-\infty}^\infty \dd x_2 \cdots \int_{-\infty}^\infty\dd x_N P^{\rm st}(x,x_2, \cdots, x_N)  \;.
\end{equation}
Since the integrand in the expression of the JPDF, for fixed $u$, is a simple product of independent Gaussians, the $(N-1)$-fold integral in Eq. (\ref{rho-x}) can be performed trivially, leading to 
\bea \label{def_rho2}
\rho(x, N) = \frac{c\,r_H}{4}  \int_0^1 \dd u \; u^{R_1 - 1} (1 - u)^{R_2 - 1} \left[ \frac{u}{\mu_2} + \frac{1 - u}{\mu_1}\right] \frac{1}{\sqrt{2 \pi V(u)}} e^{-\frac{x^2}{2 V(u)}} \;.
\eea
Evidently, the density $\rho(x,N)$ is a symmetric function of $x$ around $x=0$. 
Unfortunately this integral can not be performed explicitly. 
However, one can easily derive the asymptotic behaviors of $\rho(x, N)$ for small and large $x$. 
 
\vspace*{0.5cm}
\noindent{\it The limit $x \to 0$.} Expanding the Gaussian in (\ref{def_rho2}) up to quadratic order, one immediately finds 
\bea
\rho(x, N) \underset{x \to 0}{\longrightarrow} \sqrt{\frac{r_H}{2 D}} \left(C_1 - C_2 \frac{r_H}{2 D} x^2\right) \;,
\eea 
where $C_1, C_2$ are dimensionless constants given by
\bea
C_1 = \frac{c}{4 \sqrt{\pi}} \int_0^1 \dd u \; u^{R_1 - 1} (1 - u)^{R_2 - 1} \sqrt{u\frac{r_H}{\mu_2} + (1 - u)\frac{r_H}{\mu_1}}  \;
\eea
and
\bea
C_2 = \frac{c}{4 \sqrt{\pi} } \int_0^1 \dd u \; u^{R_1 - 1} (1 - u)^{R_2 - 1} \left(\sqrt{u\frac{r_H}{\mu_2} + (1 - u)\frac{r_H}{\mu_1}}\right)^{-1}  \;,
\eea
where $c$ and $r_H$ are given in Eq. (\ref{const_sm}).

\vspace*{0.5cm}
\noindent{\it The limit $x \to \infty$.} The $x \to +\infty$ limit is a bit more tricky. We note that the Gaussian inside the integrand, for fixed $u$, reads
$e^{-x^2/(2 V(u))}$ where $V(u)$ is given in Eq. (\ref{def_V_of_u}). Since $\mu_1 > \mu_2$, the variance $V(u)$ increases monotonically as $u$ increases from $0$ to $1$. Hence, the dominant contribution to the integral for $x \to \infty$ will clearly originate from the vicinity of $u=1$. We therefore expand $1/V(u)$ \blue{around} $u=1$. To proceed, it is convenient to first define  
\bea \label{def_psi}
\psi(u) = \frac{1}{\blue{2 V(u)}} =  \frac{1}{2 D \left( \frac{u}{\mu_2} + \frac{1 - u}{\mu_1} \right)} \;.
\eea
\blue{Hence}, to leading order for large $x$, we get
\bea
\rho(x, N) {\approx} \frac{c \, r_H}{4} e^{-x^2 \psi(1)} \int_0^1 \dd u\; u^{R_1 - 1} (1 - u)^{R_2 - 1} \left[ \frac{u}{\mu_2} + \frac{1 - u}{\mu_1} \right] \frac{1}{\sqrt{\pi \psi(1)}} e^{- x^2 (u - 1) \psi'(1)} \;.
\eea
Changing variable to $v = x^2(1-u)$, the bounds of the integral become $[0, x^2]$. \blue{Therefore} in the large $x$ limit we can approximate the bounds to be $[0, +\infty[$. Then to leading order we obtain
\bea
\rho(x, N) \approx \frac{c\, r_H}{4} x^{-2 R_2} e^{-x^2 \frac{\mu_2}{2 D}} \frac{1}{\mu_2}\left[\frac{\pi \mu_2}{2 D}\right]^{-1/2} \int_0^{+\infty} \dd v \; v^{R_2 - 1} e^{- v \frac{\mu_2}{2 D} \frac{\mu_1 - \mu_2}{\mu_1} } \;.
\eea
This integral can readily be performed and yields
\bea \label{rho-tail}
\rho(x, N) \approx  \frac{\Gamma(1 + R_1 + R_2)}{\Gamma(1 + R_1)} \frac{1}{2 \sqrt{\pi}} \frac{r_H}{r_2} \left( \frac{\mu_1}{\mu_1 - \mu_2} \right)^{R_2}  \left(\frac{\mu_2}{2 D} \right)^{-R_2-1/2} \, x^{-2 R_2}\,e^{-\frac{\mu_2}{2 D}\,x^2} \;.
\eea
Therefore, summarizing the asymptotic behaviors of $\rho(x,N)$, we have
\bea
\rho(x, N) \approx \begin{dcases}
\sqrt{\frac{r_H}{2 D}} \left(C_1 - C_2 \frac{r_H}{2 D} x^2\right) &\mbox{~~when~~} x \to 0\\
C \, x^{-2 R_2}\, e^{- \frac{\mu_2}{2 D}\,x^2} &\mbox{~~when~~} x \to +\infty
\end{dcases}\;,
\eea
where $C$ is just a constant which can be read off from Eq. (\ref{rho-tail}). 

\vspace*{0.3cm}

In Fig. \ref{fig-supp-mat} we compare the analytical prediction in Eq. (\ref{def_rho2})
with numerical simulations. These simulations were performed in two different ways: (a) by direct sampling of the JPDF in Eq. (\ref{Pst_sm}) where we draw a random number $u \in [0,1]$ distributed via $h(u)$ in Eq. (\ref{h-p}) and then draw $N$ independent Gaussian random variables each with zero mean and \blue{variance $V(u)$}. From this one then computes the average density. (b) Direct Monte-Carlo simulation of the Langevin dynamics in Eqs. (\ref{dynamic1}), (\ref{dynamic2}) and (\ref{eta}). We have checked that both are in perfect agreement with our analytic predictions. Clearly the direct sampling method is much more efficient than the Monte-Carlo simulations.

\subsection{Correlator in the NESS}

In this subsection, we derive explicitly the correlations between the positions of the particles in the NESS. 
From Eqs. (\ref{jpdf}) and (\ref{h-p}) it is clear that $\langle x_i \rangle = 0$ and $\langle x_i x_j \rangle = 0$ for $i \neq j$, because of the symmetry of the Gaussians. The first nonzero correlator turns out to be $\langle x_i^2 x_j^2 \rangle - \langle x_i^2 \rangle \langle x_j^2 \rangle$. From Eq. (\ref{jpdf}) we get
\bea
\langle x_i^2 \rangle = \int_0^1 \dd u \; h(u) \left(\prod_{k \neq i} \int_{-\infty}^{+\infty} \dd x_k \; p(x_k | u)\right) \left( \int_{-\infty}^{+\infty} \dd x_i \; x_i^2 p(x_i | u) \right) \;.
\eea
Using Eq. (\ref{h-p}) we get
\bea
\langle x_i^2 \rangle = \frac{c\, r_H}{4} \int_0^1 \dd u \; u^{R_1 - 1} (1 - u)^{R_2 - 1} \left[\frac{u}{\mu_2} + \frac{1 - u}{\mu_1}\right] \blue{V(u)} \;.
\eea
This integral can be explicitly performed and it yields
\bea \label{xi-2}
\langle x_i^2 \rangle = \frac{D}{2 \mu_1\mu_2} \frac{(r_1 + r_2)^2 + 2 r_1 \mu_1 + 2 r_2 \mu_2}{(r_1 + r_2)(1 + R_1 + R_2)} \;.
\eea
A similar computation can be done for $\langle x_i^2 x_j^2 \rangle$. From Eq. (\ref{jpdf}) we get
\bea
\langle x_i^2 x_j^2 \rangle = \int_0^1 \dd u \; h(u) \left(\prod_{k \neq i, j} \int_{-\infty}^{+\infty} \dd x_k \; p(x_k | u)\right) \left( \int_{-\infty}^{+\infty} \dd x_i \; x_i^2 p(x_i | u) \right) \left( \int_{-\infty}^{+\infty} \dd x_j \; x_j^2 p(x_j | u) \right) \;.
\eea
Using Eq. (\ref{h-p}) we get 
\bea
\langle x_i^2 x_j^2 \rangle = \frac{c \, r_H}{4} \int_0^1 \dd u \; u^{R_1 - 1} (1 - u)^{R_2 - 1} \left[\frac{u}{\mu_2} + \frac{1 - u}{\mu_1}\right] \blue{[V(u)]^2} \;,
\eea
\blue{where we recall that $V(u)$ is given in Eq.~(\ref{def_V_of_u})}. Once again we can perform this integral and we obtain
\begin{align} 
&\langle x_i^2 x_j^2 \rangle = \frac{D^2}{4 \mu_1^2 \mu_2^2} \frac{(r_1 + r_2)^3 + 6(\mu_1 r_1 + \mu_2 r_2)(r_1 + r_2) + 8 r_1 \mu_1^2 + 8 r_2 \mu_2^2 }{(r_1 + r_2)(1 + R_1 + R_2)(2 + R_1 + R_2)} \label{xi-xj}
\end{align}
Putting Eq. (\ref{xi-2}) and Eq. (\ref{xi-xj}) together we obtain the final expression for the first non-trivial correlator for $i \neq j$
\bea \label{correl_sm}
\langle x_i^2 x_j^2 \rangle - \langle x_i^2 \rangle \langle x_j^2 \rangle = \frac{D^2}{4 \mu_1^2 \mu_2^2} \frac{2 r_1 r_2 (\mu_1 - \mu_2)^2 (( r_1 + r_2 )^2 + 8 \mu_1 \mu_2 (1 + R_1 + R_2) + 2 r_1 \mu_1 + 2 r_2 \mu_2)}{(r_1 + r_2)^2 (1 + R_1 + R_2)^2 (4 \mu_1 \mu_2 + r_1 \mu_2 + r_2 \mu_1)} \;.
\eea
The fact that the right hand side of Eq. (\ref{correl_sm}) is positive indicates that the positions of the particles are positively correlated. These positive correlations emerge from the effective attraction between the particles generated by the simultaneous switching of the background harmonic potential. Note that for $r_1=r_2=r$, we recover the formula in Eq. (14) in the main text. Besides, when $\mu_1 = \mu_2$, these correlations vanish as expected since the particles remain independent at all times when the trap is static.

\subsection{Order statistics}\label{sec:order}

\begin{figure}[t]
\includegraphics[width=0.7\linewidth]{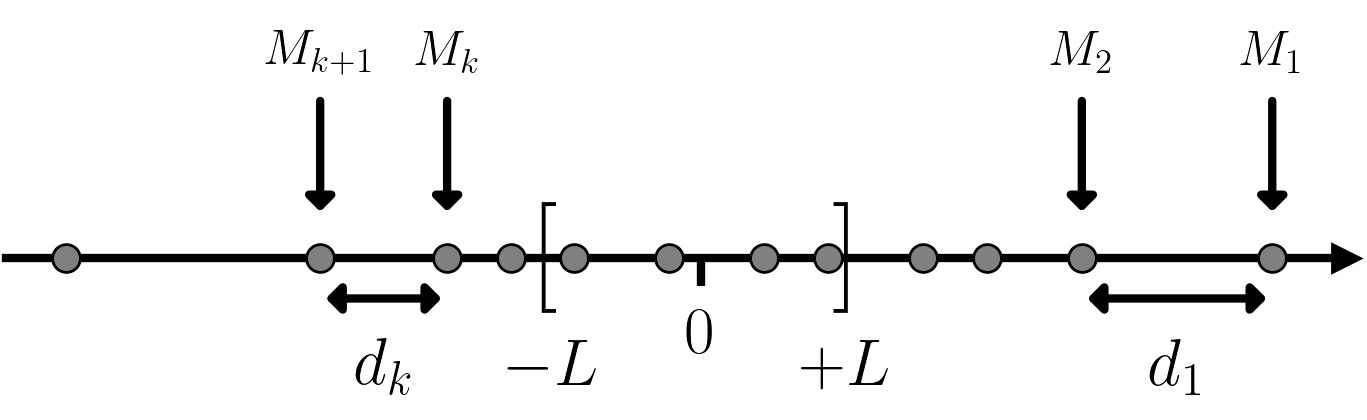}
\caption{Schematic representation of the correlated gas. The grey dots represent the positions of the particles and $M_1>M_2>~\cdots$ represent the ordered positions from right to left. The gap between the positions of the $k$-th and $(k+1)$-th particles is denoted by $d_k = M_{k}-M_{k+1}$. In this particular sample the number of particles in the interval $[-L,+L]$ is $N_L = 4$.}\label{fig_order}
\end{figure}

In this subsection, we study the order statistics by sorting the positions $\{x_1, \cdots, x_N\}$ in decreasing order $\{M_1 \geq M_2 \geq \cdots \geq M_N\}$. Then $M_k$ corresponds to the position of the $k$-th particle counting from the right (see Fig. \ref{fig_order}). We recall that $\mu_1 > \mu_2$, 
without any loss of generality. To compute the statistics of $M_k$, we note from the structure of the JPDF in Eq. (\ref{jpdf}) that we need to compute the distribution of $M_k$ for $N$ i.i.d. random Gaussian variables, each with zero mean and variance $V(u)$ given in Eq.~\ref{def_V_of_u} for a fixed $u$ and then average over $u$ drawn from $h(u)$ in Eq. (\ref{h-p}). This procedure was worked out in detail in  
Ref. \cite{BLMS_23-2} for general conditionally independent identical variables. Here we do not repeat the detailed derivation but just outline the main steps involved. First we will set $k = \alpha N$ with $0<\alpha<1$. By setting $\alpha = O(1)$, we can probe the order statistics deep inside the bulk of the gas, while by setting $\alpha = O(1/N)$ we can probe the order statistics at the edge, e.g., the statistics of $M_1, M_2$, etc. Below we start with the bulk with $\alpha = O(1)$ and later recover the edge results by taking the $\alpha \to 0$ limit.  

The first step to compute the order statistics in the bulk is to define the $\alpha$-quantile as
\bea \label{q-def}
\alpha = \int_{q(\alpha, u)}^{+\infty} p(x | u) \dd x\;,
\eea
where $p(x \vert u)$ is a simple Gaussian given in Eq. (\ref{h-p}). This gives explicitly
\bea \label{q}
q(\alpha, u) =  \sqrt{2\,V(u)} \, \blue{\; {\rm erfc}^{-1}\, \left(2\alpha\right)} \;,
\eea
where ${\rm erfc}^{-1}(z)$ is the inverse of the complementary error function ${\rm erfc}(z) = (2/\sqrt{\pi}) \int_z^{+\infty} e^{-u^2} \dd u$.
In terms of the $\alpha$-quantile, the PDF of $M_k$ can then be expressed as \cite{BLMS_23-2}
\bea \label{mk-delta}
{\rm Prob.}[M_k = w] = \int_0^1 \dd u \; h(u) \delta( q(\alpha, u) - w ) \;.
\eea
For compactness, let us denote 
\bea \label{def_beta}
\beta = {\rm erfc}^{-1}(2\alpha) \;.
\eea
Substituting Eq. (\ref{q}) in Eq. (\ref{mk-delta}) we obtain
\bea \label{Mk-explicit}
{\rm Prob.}[M_k = w] = \int_0^1 \dd u \; h(u) \; \delta\left( \beta \sqrt{\frac{2 D}{\mu_1 \mu_2} (u \mu_1 + (1 - u) \mu_2)} - w\right) \;.
\eea
This integral can be performed leading to the scaling form
\bea \label{Mk}
{\rm Prob.}[M_k = w] = \sqrt{\frac{r_H}{4 D \beta^2}} \; f\left(w \sqrt{\frac{r_H}{4 D \beta^2}}\right) \;,
\eea
where we recall that $r_H = \frac{2}{\frac{1}{r_1} + \frac{1}{r_2}}$ is the harmonic mean of the switching rates. The scaling function $f(z)$ is supported over the finite interval $\sqrt{R_{H,1}} < z < \sqrt{R_{H,2}}$ and is given explicitly by
\bea \label{f-z}
f(z) = c \frac{R_{H,1}^{R_1 - 1} R_{H,2}^{R_2 - 1}}{(R_{H,2} - R_{H,1})^{R_1 + R_2 - 1}}  |z|^3 \left(1 - \frac{z^2}{R_{H, 2}}\right)^{R_2 - 1} \left(\frac{z^2}{R_{H, 1}} - 1 \right)^{R_1 - 1}   \;,
\eea
where
\bea
R_{H, 1} = \frac{r_H}{2 \mu_1} \mbox{~~and~~} R_{H, 2} = \frac{r_H}{2 \mu_2} \;.
\eea
One can easily verify the normalization 
\bea \label{norm}
\int_{\sqrt{R_{H,1}}}^{\sqrt{R_{H,2}}} \dd z \, f(z) = 1 \;.
\eea
For $r_1 = r_2 = r$, this reproduces the result in Eq. (16) in the main text. 

As discussed in the main text, the fact that the scaling function for the $k$-th maximum is supported over a finite interval is rather unusual since
in most known examples \blue{\cite{MPS_20}}, the associated scaling function of $M_k$ has an infinite (or semi-infinite) support. Moreover, the shape of this scaling function can be tuned by varying the parameters $r_1, r_2, \mu_1$ and $\mu_2$. For instance, from Eq. (\ref{f-z}), if $R_1 >1$ and $R_2>1$, the scaling function $f(z)$ vanishes at both edges of the support (see the left panel of Fig. 2 in the main text). If $R_1<1$ and $R_2>1$, the scaling function diverges at the lower edge but vanishes at the upper edge (see the middle panel of Fig. 2 in the main text). Similarly, if $R_1<1$ and $R_2<1$, the scaling function diverges at both edges (see the right panel of Fig. 2 in the main text).

\vspace{0.5cm}

Since the scaling function $f(z)$, given in Eq. (\ref{f-z}), is independent \blue{of $\alpha$}, it also holds for $M_k$ when $k =O(1)$, i.e., $\alpha = O(1/N)$. The only difference is in the scale factor in Eq. (\ref{Mk}). Indeed, by setting $\alpha = k/N$, with $k=O(1)$, 
one finds from Eq.~(\ref{def_beta}) to leading order for large $N$ 
\bea \label{asympt_beta}
\beta = {\rm erfc}^{-1}[2 \alpha] \approx \sqrt{ \ln N} \;,
\eea
independent of $k$. Hence, for all $k=O(1)$, we have
\bea \label{M1}
{\rm Prob.}[M_k = w] = \sqrt{\frac{r_H}{4 D \ln N}} \; f\left(w \sqrt{\frac{r_H}{4 D \ln N}} \right) \;,
\eea
where the scaling function $f(z)$ is given in Eq. (\ref{f-z}). Thus the scaling function $f(z)$ is universal, i.e., independent of the order $k$, either in the bulk or at the edges.  

\subsection{Gap statistics}

In this subsection, we compute the statistics of the gap $d_k = M_{k} - M_{k+1}$. Once again, we will exploit the conditionally i.i.d. structure of the joint distribution in Eq. (\ref{jpdf}) and follow the general procedure outlined in Refs.~\cite{BLMS_23,BLMS_23-2}. For $N$ i.i.d. variables distributed via $p(x\vert u)$ in Eq. (\ref{h-p}) with $u$ fixed, the gap $g$ is distributed in the large $N$ limit as $N p[q(\alpha, u) | u] e^{- N p[q(\alpha, u)] g}$ \cite{BLMS_23,BLMS_23-2}, where $q(\alpha, u)$ is the $\alpha$-quantile defined in Eq. (\ref{q}). Averaging over $u$, drawn from $h(u)$ in Eq. (\ref{h-p}), we get 
\bea \label{dk-orig}
{\rm Prob.}[d_k = g] = \int_0^{1} \dd u \; h(u) \; N p[q(\alpha, u) | u] e^{- N p[q(\alpha, u)] g} \;.
\eea
Using Eq. (\ref{h-p}) and Eq. (\ref{q}) we can re-write Eq. (\ref{dk-orig}) in a scaling form
\bea \label{dk}
{\rm Prob.}[d_k = g] = N\,\sqrt{\frac{\mu_H}{4 \pi D}} \,e^{-\beta^2}\, 
F\left( \sqrt{\frac{\mu_H}{4 \pi D}} e^{-\beta^2} N\, g \right) \;,
\eea
where $\mu_H = 2 \mu_1 \mu_2/(\mu_1+\mu_2)$, the constant $\beta$ is given in Eq. (\ref{def_beta}) and the scaling function $F(z)$, supported on $z \geq  0$, is given by
\begin{align} 
F(z) = c \frac{r_H}{4} \int_0^1 \dd u\; u^{R_1-1} (1 - u)^{R_2-1} \left[ \frac{1 - u}{\mu_1} + \frac{u}{\mu_2} \right] \; \sqrt{\frac{\mu_1 + \mu_2}{u \mu_1 + (1 - u)\mu_2}} \exp( - z \sqrt{\frac{\mu_1 + \mu_2}{u \mu_1 + (1 - u)\mu_2}} ) \;. \label{G}
\end{align}
One can check that $F(z)$ is normalized to 1, i.e., $\int_0^\infty \dd z \, F(z) = 1$. While we could not compute this integral explicitly, the asymptotic behavior of $F(z)$ can be easily extracted from Eq. (\ref{G}).  

\begin{figure*}
\centering
\includegraphics[width=0.32\textwidth]{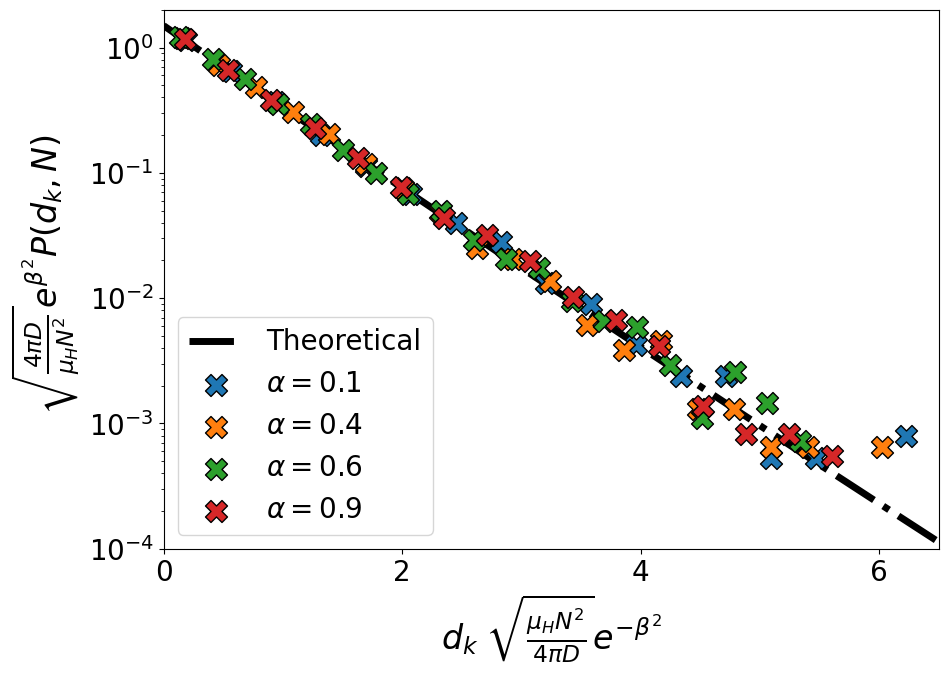} 
\hfill
\includegraphics[width=0.32\textwidth]{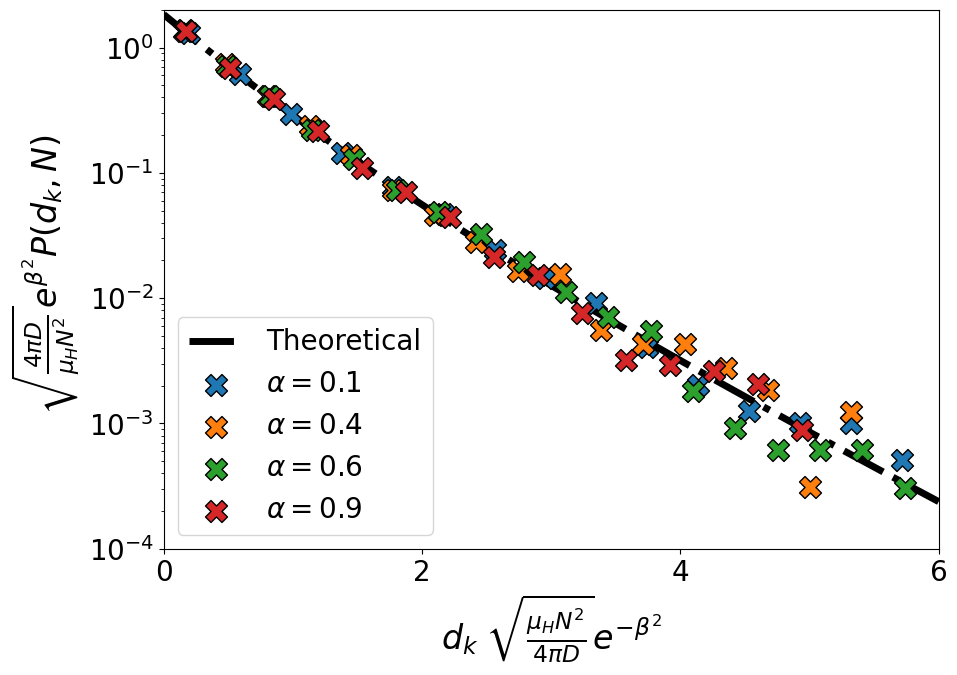} 
\hfill
\includegraphics[width=0.32\textwidth]{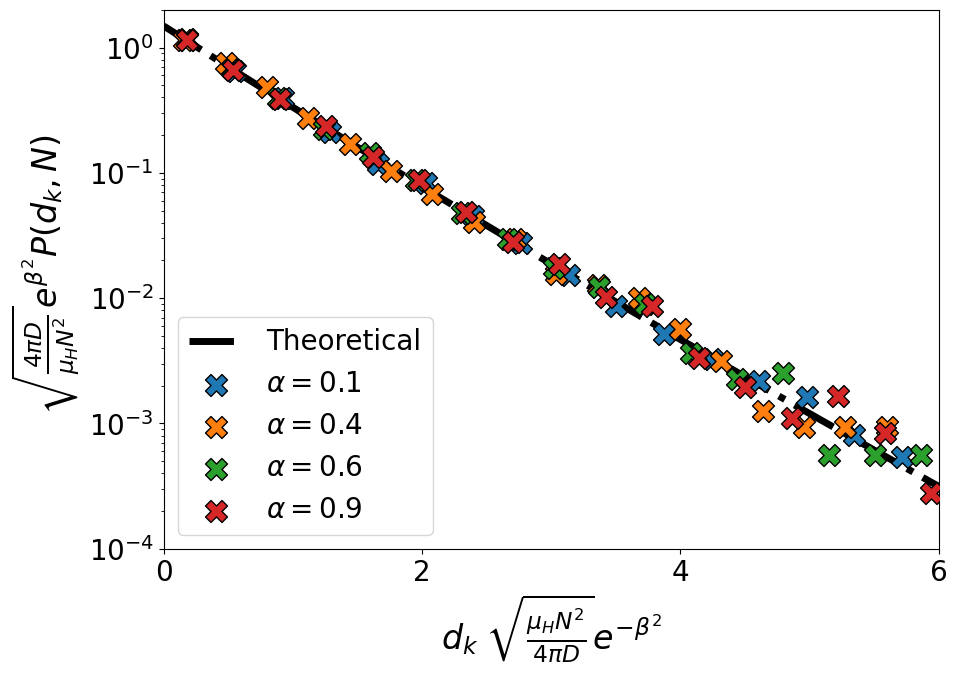} 
\caption{Scaling collapse of the distribution of the $k$-th gap as in Eq.~(\ref{dk}) for different values of $\alpha = k/N$ and different values of the parameters. We set $r_1 = r_2 = 1$, $D=1$, $N = 10^{6}$ and vary $\mu_1$ and $\mu_2$. From left to right we used respectively $\mu_1 = 0.4, \, \mu_2 = 0.2$ then $\mu_1 = 2, \, \mu_2 = 0.4$ and finally $\mu_1 = 2,\, \mu_2 = 1$. The dashed black line corresponds to the theoretical prediction given in Eq. (\ref{G}) and the symbols are the numerical results. Different colors correspond to different values of $\alpha$. The numerical results were obtained by sampling $10^{5}$ examples directly from the NESS distribution given in Eq. (\ref{jpdf}).} \label{fig:gap}
\end{figure*}

\vspace*{0.5cm}
\noindent{\it The limit $z \to 0$.} In this limit, expanding $e^{-s z} \sim 1 - s z$, we get
\bea \label{G-small-z}
F(z) \approx  B_1 - B_2 z \;,
\eea
where the constants $B_1$ and $B_2$ are given by
\bea \label{A1}
B_1 = \frac{c\, r_H}{4} \int_0^{1}\dd u\; u^{R_1 - 1} (1 - u)^{R_2 - 1} \left[\frac{1-u}{\mu_1} + \frac{u}{\mu_2}\right] \sqrt{\frac{\mu_1 + \mu_2}{u \mu_1 + (1 - u) \mu_2}} \;,
\eea
and 
\bea \label{A2}
B_2 = \frac{c\, r_H}{4} \int_0^{1}\dd u\; u^{R_1 - 1} (1 - u)^{R_2 - 1} \left[ \frac{1 - u}{\mu_1} + \frac{u}{\mu_2} \right] \frac{\mu_1 + \mu_2}{u \mu_1 + (1 - u) \mu_2} = \frac{r_H}{\mu_H} \frac{r_2 \mu_1 + r_1 \mu_2}{r_1 r_2}\;.
\eea

\vspace*{0.5cm}
\noindent{\it The limit $z \to \infty$.} Since we have set $\mu_1 > \mu_2$, we find that the function 
\bea \label{def_phi2}
\phi(u) = \sqrt{\frac{\mu_1 + \mu_2}{u \mu_1 + (1 - u)\mu_2}} 
\eea
that appears inside the argument of the exponential in Eq. (\ref{G}) is a monotonically decreasing function of $u$ for $u \in [0,1]$. Consequently, the dominant contribution to the integral for large $z$ comes from the vicinity of $u=1$. Expanding $\phi(u)$ near $u=1$, we get for large $z$
\bea
F(z) \approx \frac{c\, r_H}{4} \,e^{-z \phi(1)}\, \phi(1) \, \int_0^1 \dd u \; u^{R_1 - 1}(1 - u)^{R_2 -1} \left[ \frac{1 - u}{\mu_1} + \frac{u}{\mu_2} \right]  e^{- z (u-1)\phi'(1)} \;.
\eea
Using Eq. (\ref{def_phi2}) gives 
\bea
F(z) \approx \frac{c\, r_H}{4} \sqrt{\frac{\mu_1+\mu_2}{\mu_1}}e^{-z \sqrt{\frac{\mu_1 + \mu_2}{\mu_1}}} \int_0^1 \dd u \; u^{R_1 - 1} (1 - u)^{R_2 - 1} \left[ \frac{1 - u}{\mu_1} + \frac{u}{\mu_2} \right] e^{-z (1 - u) \frac{\sqrt{\mu_1 + \mu_2}}{2\mu_1^{3/2}}(\mu_1 - \mu_2) } \;.
\eea
Changing variable to $v = (1 - u) z$ we get, for large $z$,
\bea
F(z) \approx \frac{c \, r_H}{4} z^{-R_2} \frac{1}{\mu_2} \sqrt{\frac{\mu_1 + \mu_2}{\mu_1}} e^{-z\sqrt{\frac{\mu_1 + \mu_2}{\mu_1}}} \int_0^{+\infty} \dd v \; v^{R_2 - 1} e^{-v \frac{\sqrt{\mu_1 + \mu_2}}{2 \mu_1^{3/2}} (\mu_1 - \mu_2)} \;.
\eea
This integral over $v$ can be done explicitly, leading to 
\bea \label{G-large-z}
F(z) \approx B \, \frac{e^{-z \sqrt{\frac{\mu_1 + \mu_2}{\mu_1}}}}{z^{R_2}} \qquad {\rm where} \qquad B = \frac{\Gamma(1 + R_1 + R_2)}{\Gamma(1 + R_1)} \frac{r_H}{r_2} \left(\frac{\mu_1}{\mu_1 - \mu_2}\right)^{R_2} \left(\sqrt{\frac{4 \mu_1}{\mu_1 + \mu_2}}\right)^{R_2 - 1}  \;.
\eea
To summarize, the asymptotics of $F(z)$ are given by
\bea \label{F-tails}
F(z) \longrightarrow \begin{cases}
B_1 - B_2 z \;  &\mbox{~~for~~} z \ll 1\\
B \, z^{-R_2} e^{-z \sqrt{\frac{\mu_1 + \mu_2}{\mu_1}}}  &\mbox{~~for~~} z \gg 1
\end{cases} \;.
\eea
In Fig. \ref{fig:gap}, we compare this analytical scaling function $F(z)$ in Eq. (\ref{G}) with numerical simulations, showing an excellent agreement. 

\subsection{Full counting statistics (FCS)}

\begin{figure*}
\centering
\includegraphics[width=0.32\textwidth]{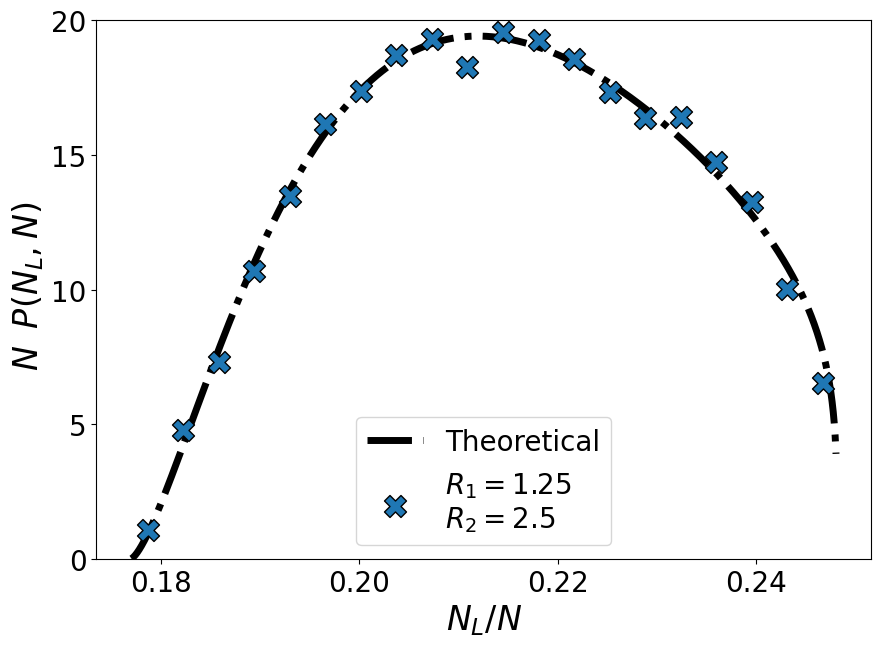} 
\hfill
\includegraphics[width=0.32\textwidth]{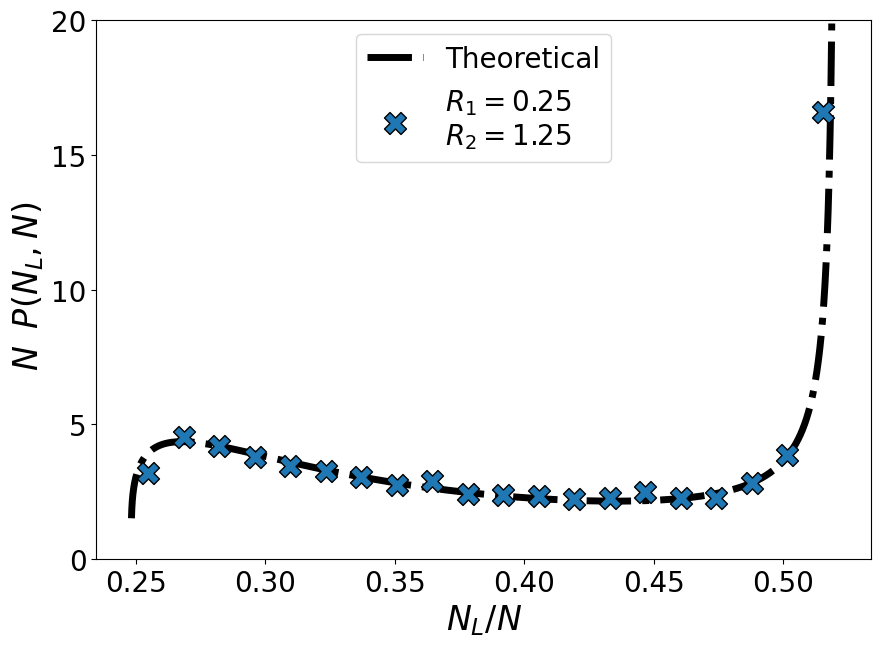} 
\hfill
\includegraphics[width=0.32\textwidth]{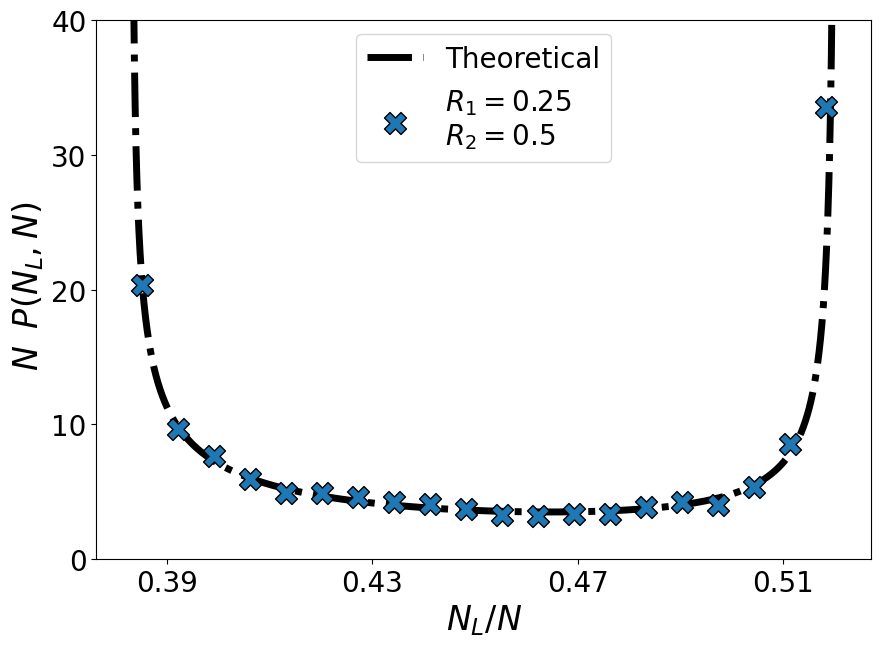} 
\caption{Scaling collapse of the distribution of the number of particles $N_L$ in $[-L,L]$ as in Eq.~(\ref{NL}) for different values of the parameters $R_1$ and $R_2$. We set $L=0.5, r_1 = r_2 = 1$, $D=1$, $N = 10^{6}$ and vary $\mu_1$ and $\mu_2$. From left to right we used respectively $\mu_1 = 0.4, \, \mu_2 = 0.2$ then $\mu_1 = 2, \, \mu_2 = 0.4$ and finally $\mu_1 = 2,\, \mu_2 = 1$. The dashed black line corresponds to the theoretical prediction given in Eq. (\ref{H}) and the symbols are the numerical results. The numerical results were obtained by sampling $10^{5}$ examples directly from the NESS distribution given in Eq. (\ref{jpdf}).} \label{fig_fcs}
\end{figure*}

Finally, we compute the FCS, i.e., the distribution of the number $N_L$ of particles inside the interval $[-L, L]$ around the origin. Exploiting again the conditionally i.i.d. structure of the JPDF in Eq. (\ref{jpdf}) and adapting the formalism in Refs. \cite{BLMS_23,BLMS_23-2} we get for the probability distribution of $N_L$
\bea
P(N_L, N) = \frac{1}{N} \int_0^1 \dd u\; h(u) \; \delta\left[ \frac{N_L}{N} - \int_{-L}^L \dd x \; p(x | u) \right] \;.
\eea
where $p(x | u)$ and $h(u)$ are given in Eq. (\ref{h-p}). Using Eq. (\ref{h-p}) we can express the distribution in a scaling form as
\bea \label{NL}
P(N_L,N) \approx \frac{1}{N} H\left( \frac{N_L}{N} \right) \;,
\eea
where the scaling function $H(z)$ is supported over ${\rm erf}(\sqrt{\gamma/R_{H, 2}}) < z < {\rm erf}(\sqrt{\gamma/R_{H, 1}})$. Here we have denoted 
\bea \label{gamma}
\gamma= \frac{r_H L^2}{4 D} \;.
\eea
The scaling function $H(z)$ is given explicitly by
\begin{align}
H(z) = \frac{c \, \, \gamma^2}{(R_{H, 2} - R_{H, 1})^{R_1 + R_2 - 1}} \frac{\sqrt{\pi}}{2} e^{u(z)^2} \frac{1}{u(z)^5} \left( \frac{\gamma}{u(z)^2} - R_{H, 1} \right)^{R_1 - 1} \left(R_{H, 2} - \frac{\gamma}{u(z)^2}\right)^{R_2 - 1} \;, \label{H}
\end{align}
where \blue{$u(z) = {\rm erf}^{-1}(z)$ is} the inverse error function. In contrast to the results obtained in Ref. \cite{BLMS_23}, we can see that for this system the FCS have a richer variety of behaviors with a finite support contained in $[0, 1]$ and possible divergences at the edges of the support. 
One can check that this scaling function $H(z)$ is normalized to unity over its support ${\rm erf}(\sqrt{\gamma/R_{H, 2}}) < z < {\rm erf}(\sqrt{\gamma/R_{H, 1}})$. In Fig. \ref{fig_fcs}, we compare this analytical scaling function $H(z)$ in Eq. (\ref{H}) with the numerically obtained scaling function, and find an excellent agreement.

We remark on an interesting fact. Since $N_L \in [0, N]$, the scaling variable $z=N_L/N$ has an allowed range $z \in [0,1]$. However, we find that, in the limit $N \to \infty$, the scaling function $H(z)$ is supported over a smaller interval  $z \in [{\rm erf}(\sqrt{\gamma/R_{H, 2}}), {\rm erf}(\sqrt{\gamma/R_{H, 1}})] \subset [0,1]$. Hence, the probability of having $N_L < N {\rm erf}(\sqrt{\gamma/R_{H, 2}})$ or $N_L > N {\rm erf}(\sqrt{\gamma/R_{H, 1}})$ is vanishingly small in the large $N$ limit. It would be interesting to investigate the leading large $N$ behavior of this vanishing probability outside this shorter range.

\section{Resetting limit}

As stated in the Letter, if we take the limit $\mu_1 \to +\infty, r_1 \to +\infty$ and $\mu_2 \to 0, r_2 \to r$ we should recover the resetting model we previously studied in \cite{BLMS_23}. \blue{This is because in the limit $\mu_1 \to \infty$, the potential is extremely steep in phase 1, and hence the particle returns back instantaneously to the origin. In addition, when $r_1 \to \infty$, it gets reset to the origin with rate $r_2=r$, but does not spend any finite time at the origin (see Fig. 1 in the main text). This is precisely the limit that was studied in Ref. \cite{BLMS_23}.} In order to recover this limit, we start with the joint distribution 
\bea \label{jpdf_2}
P^{\rm st}(\vec{x}) = \int_0^1 \dd u\; h(u) \prod_{k = 1}^N p(x_k | u) \;,
\eea
where 
\bea \label{h-p_2}
h(u) = \frac{c \, r_H}{4} u^{R_1 - 1} (1 - u)^{R_2 - 1} \left[\frac{u}{\mu_2} + \frac{1 - u}{\mu_1} \right] \mbox{~~and~~} p(x | u) = \sqrt{\frac{1}{2 \pi V(u)}} \exp( - \frac{x^2}{2 V(u)} )  \;,
\eea
where $V(u)$ is given in Eq. (\ref{def_V_of_u}) and the constants are 
\bea \label{const_sm_2}
c = \frac{\Gamma(R_1 + R_2 + 1)}{\Gamma(R_1 + 1)\Gamma(R_2 + 1)}\;, \quad R_1 = \frac{r_1}{2 \mu_1} \;, \quad R_2 = \frac{r_2}{2 \mu_2} \quad {\rm and} \quad r_H = 2 \frac{r_1 r_2}{r_1+r_2} \;.
\eea
Let us first consider the limit $\mu_1 \to \infty$ and $\mu_2 \to 0$ with $r_1$ and $r_2$ fixed. In this limit, $R_1 \to 0$, $R_2 \to \infty$ and $h(u)$ becomes
\bea \label{h_limit}
h(u) = \frac{r_2}{r_1+r_2} \delta(u) + R_2 \,\tilde h \left( R_2 u \right) \qquad {\rm where} \quad \tilde h(v) = \frac{r_1}{r_1+r_2} \, e^{-v} \;.
\eea
Evidently $\int_0^1 \dd u h(u) = 1$, as it should be. Now if one takes the $r_1 \to \infty$ limit, the delta function disappears and one recovers a purely exponential function $h(u)$ -- rescaled by $R_2$. We now consider the second factor $p(x\vert u)$ in Eq. (\ref{h-p_2}). In this limit one has
\bea \label{gauss_lim}
p(x\vert u) \longrightarrow \sqrt{\frac{\mu_2}{2 \pi D u}} e^{-\frac{\mu_2 x^2}{2 D u}} \;.
\eea
Hence, in the limits $\mu_2 \to 0$, $\mu_1 \to \infty$ and $r_1 \to \infty$, we get the limiting joint distribution
\bea \label{jpdf_lim}
P^{\rm st}(\vec{x})   \longrightarrow R_2\, \int_0^1 \dd u   \, e^{- u R_2} \, \prod_{i=1}^N \sqrt{\frac{\mu_2}{2 \pi D u}} e^{-\frac{\mu_2 x_i^2}{2 D u}} \;.
\eea
Making the change of variable $u = 2 \mu_2 \, \tau$, and denoting $r_2=r$, we get 
\bea \label{jpdf_final}
P^{\rm st}(\vec{x}) \longrightarrow \int_0^{+\infty} \dd \tau \; r e^{-r\tau} \prod_{i = 1}^{N} \frac{1}{\sqrt{4 \pi D \tau}} \exp(-\frac{x_i^2}{4 D \tau}) \;,
\eea
thus recovering the result of Ref. \cite{BLMS_23}. In summary, to recover the limit of instantaneous resetting studied in Ref. \cite{BLMS_23}, the proper limits are rather subtle and are given by    
\bea \label{limits}
\begin{cases}
\mu_1 \to +\infty\\
r_1 \to +\infty\\
\end{cases} \mbox{~~and~~} \begin{cases} \mu_2 \to 0 \\
r_2 \to r \end{cases} \mbox{~~such that~~} \begin{cases}
R_1 = \frac{r_1}{2 \mu_1} \to 0 \\
R_2 = \frac{r_2}{2 \mu_2} \to +\infty
\end{cases}\;.
\eea
Note that the $\mu_1 \to \infty$ limit is taken before the $r_1 \to \infty$ limit, such that the ratio $R_1 = r_1/(2 \mu_1) \to 0$.

\section{Extension to higher spatial dimensions}

In this section, we show that our results can be straightforwardly extended to study $N$ non-interacting diffusing particles in $d$ dimensions and in the presence of a switching isotropic $d$ dimensional harmonic trap. The derivation presented in Section \ref{sec_deriv} can be generalised to higher dimensions and the joint distribution of the positions of the particles ${\bf x}_1, \cdots, {\bf x}_N$ (where ${\bf x}_i$ is now a $d$-dimensional vector) reads
\bea \label{jpdf_d}
P^{\rm st}({\bf x}_1, \cdots, {\bf x}_N) = \int_0^1 \dd u\; h(u) \prod_{k = 1}^N p({\bf x}_k | u) \;,
\eea
where 
\begin{equation} \label{h-p_d}
h(u) = \frac{c \, r_H}{4} u^{R_1 - 1} (1 - u)^{R_2 - 1} \left[\frac{u}{\mu_2} + \frac{1 - u}{\mu_1} \right] \mbox{~~and~~} p({\bf x} | u) = \left(\frac{1}{2 \pi V(u)}\right)^{d/2} \exp( - \frac{{z}^2}{2 V(u)} )  \;,
\end{equation}
where $V(u)$ is given in Eq. (\ref{def_V_of_u}), $z = ||{\bf x}||$ is the distance from the origin and the constants are 
\bea \label{const_sm_d}
c = \frac{\Gamma(R_1 + R_2 + 1)}{\Gamma(R_1 + 1)\Gamma(R_2 + 1)}\;, \quad R_1 = \frac{r_1}{2 \mu_1} \;, \quad R_2 = \frac{r_2}{2 \mu_2} \quad {\rm and} \quad r_H = 2 \frac{r_1 r_2}{r_1+r_2} \;.
\eea
Note the difference between the $d$-dimensional case and the one-dimensional case is that here there are actually $d\, N$ Gaussian factors in the product since $\exp(- a z^2)= \exp{-a [ (x^{(1)})^2 + \cdots + (x^{(d)})^2 ] }$ where $x^{(i)}$ denotes the $i$-th spatial component of ${\bf x}$.

We consider $N$ random variables $\{z_i \}$ where $z_i$ denotes the radial distance of the $i$-th particle from the centre of the trap. We sort these radii in decreasing order and denote them by 
$M_1~\geq~M_2~\geq~\cdots~\geq~M_N$ where 
\bea \label{def-M1-z}
M_1 = \max_{1 \leq i \leq N} z_i \;.
\eea
As in the one dimensional case, we will set $k = \alpha N $ in $M_k$. When $\alpha = O(1)$ this gives the statistics of the position of a particle deep inside the bulk, while when $\alpha \sim O(1/N)$, it probes the positions of the particles that are at the outer edge of the gas. 

To proceed, we first compute the JPDF of the $z_i$'s from Eq. (\ref{jpdf_d}) by moving to hyper-spherical coordinates. 
Notice that $p({\bf x} | u)$ in Eq. (\ref{h-p_d}) is spherically symmetric in $\bf{x}$. Hence
\bea\label{jpdf-z}
P^{\rm st}(z_1, \cdots, z_N) = \int_0^1 \dd u \; h(u) \prod_{i = 1}^N p(z_i | u) \;,
\eea
where the new PDF $p(z | u)$, supported on $0 < z < +\infty$, is given by
\bea \label{p-z-u}
p(z | u) = \frac{2}{\Gamma(d/2)}\frac{z^{d - 1}}{\left[ 2 V(u) \right]^{d/2}} \exp(- \frac{z^2}{2 V(u)}) \;.
\eea
The $z^{d-1}$ term comes from the integration over the angular coordinates. Note that $p(z | u)$ is normalized as $\int_0^{\infty} p(z | u)\,dz =1$. 

As in the one-dimensional case in Section \ref{sec:order}, we first compute the $\alpha$-quantile $q(\alpha,u)$, i.e., the position above which the fraction of particles is  $\alpha$. We then get 
\bea \label{def-q-z}
\alpha = \int_{q(\alpha, u)}^{+\infty} p(z | u) \dd z \;,
\eea
where $p(z | u)$ is given in Eq. (\ref{p-z-u}). \blue{Performing the change of variable $y = z^2/(2 V(u))$ and using Eq. (\ref{p-z-u}), one gets
\bea\label{q-alpha-1}
\int_{\beta^2}^{\infty} e^{-y}\, y^{d/2-1} \, dy = \Gamma(d/2)\,\alpha\, \quad {\rm where} \quad \beta = \frac{q(\alpha,u)}{\sqrt{2V(u)}} \;.
\eea}
The first relation in (\ref{q-alpha-1}) can be rewritten as 
\bea \label{implicit-q}
\Gamma\left(\frac{d}{2}, \; \frac{q^2(\alpha, u)}{2 V(u) }\right) = \Gamma(d/2) \alpha \;,
\eea
where $\Gamma(a,\, z) = \int_z^{+\infty} y^{a-1} e^{-y} \dd y$ is the incomplete Gamma function. We now define the inverse of the incomplete Gamma function with respect to the second argument as 
\bea \label{inv-inc-gamma}
\Gamma^{-1}\left[a, \; \Gamma\left(a, z\right)\right] \equiv z \;.
\eea
Consequently,  Eq.~(\ref{implicit-q}) gives 
\bea \label{explicit-q}
q(\alpha, u) = \sqrt{2  V(u)}    \; \,\sqrt{\Gamma^{-1}\left[ \frac{d}{2}, \; \Gamma(d/2) \alpha \right]} \;.
\eea
Comparing this with the one-dimensional analogue in Eq. (\ref{q}), we note that the expression for the quantile in higher dimension is identical to the one-dimensional case, except that the factor $\beta$ in Eq. (\ref{q}) gets replaced by 
\bea \label{re-def-beta}
\beta = \sqrt{\Gamma^{-1}\left[\frac{d}{2}, \; \Gamma(d/2) \alpha \right]}\;.
\eea
Consequently, the result derived for the order statistics in the one dimensional case in Eq. (\ref{Mk}) holds, i.e., 
\bea \label{MK-HD}
{\rm Prob.}[M_k = w] = \sqrt{\frac{r_H}{4 D \beta^2}} \; f\left( w \sqrt{\frac{r_H}{4 D \beta^2}} \right) \;,
\eea
where $\beta$ is given in Eq. (\ref{re-def-beta}) and the scaling function $f(z)$ is exactly the same as in the one dimensional case, namely the one in Eq (\ref{f-z}). This shows that the scaling function $f(z)$ describing the order statistics in the bulk is independent of dimension $d$, as announced in the main text. 

We now probe the outer edge of the gas by setting $\alpha = k/N$ where $k = O(1)$. For small $\alpha$, it is evident from Eq.~(\ref{q-alpha-1}) that $\beta$ is large. Consequently, to leading order for large $\beta$ (equivalently for large $N$), one gets 
\bea\label{large-beta}
\beta^{d-2} \, e^{-\beta^2} \approx \Gamma(d/2)\, \frac{k}{N} \, \quad \rm{implying} \quad \beta \approx \sqrt{\ln(N)}\, .
\eea
Substituting $\beta \approx \sqrt{\ln N}$ in Eq (MK-HD) gives for the PDF of the ordered radii at the outer edge of the gas, 
\bea\label{prob-Ln}
{\rm Prob.}[M_k = w] = \sqrt{\frac{r_H}{4 D \ln N}} \; f\left(w\sqrt{\frac{r_H}{4 D \ln N}} \right) \, ,
\eea
where $f(z)$ is the same scaling function as in Eq (\ref{f-z}). Thus the scaling function $f(z)$ is ``super-universal" in the sense that it is independent of the order $k$ and the spatial dimension $d$.